\documentclass[a4paper,fleqn,usenatbib]{mnras}

\usepackage{newtxtext,newtxmath}

\usepackage[T1]{fontenc}
\usepackage{ae,aecompl}

\usepackage{graphicx}	
\usepackage{amsmath}	
\usepackage{amssymb}	
\usepackage{todonotes}
\usepackage{bm}
\usepackage{subfig}
\usepackage{courier}
\usepackage[T1]{fontenc}
\usepackage{aecompl}
\usepackage{verbatim}
\usepackage{scrextend}
\usepackage{color}
\usepackage[english]{babel}
\usepackage{blindtext}
\usepackage{caption}
\usepackage{natbib}

\title[Shaping HR8799's outer dust belt]{Shaping HR8799's outer dust belt with an unseen planet}

\author[M. J. Read et al.]{
	M. J. Read$^{1},$\thanks{E-mail: mjr201@ast.cam.ac.uk}
	M. C. Wyatt$^{1}$, 
	S. Marino $^{1}$
	and G. M. Kennedy$^{2}$
	\\
	$^{1}$Institute of Astronomy, University of Cambridge, Madingley Road, Cambridge CB3 0HA \\
	$^{2}$Department of Physics, University of Warwick, Gibbet Hill Road, Warwick CV4 7AL
}

\date{Accepted XXX. Received YYY; in original form ZZZ}

\pubyear{2017}

\begin{document}
	\label{firstpage}
	\pagerange{\pageref{firstpage}--\pageref{lastpage}}
	\maketitle
	
	\begin{abstract}
		HR8799 is a benchmark system for direct imaging studies. It hosts two debris belts, which lie internally and externally to four giant planets. This paper considers how the four known planets and a possible fifth planet, interact with the external population of debris through N-body simulations. We find that when only the known planets are included, the inner edge of the outer belt predicted by our simulations is much closer to the outermost planet than recent ALMA observations suggest. We subsequently include a fifth planet in our simulations with a range of masses and semi-major axes, which is external to the outermost known planet. We find that a fifth planet with a mass and semi-major axis of 0.1$\mathrm{M_J}$ and 138au predicts an outer belt that agrees well with ALMA observations, whilst remaining stable for the lifetime of HR8799 and lying below current direct imaging detection thresholds. We also consider whether inward scattering of material from the outer belt can input a significant amount of mass into the inner belt. We find that for the current age of HR8799, only $\sim$1\% of the mass loss rate of the inner disk can be replenished by inward scattering. However we find that the higher rate of inward scattering during the first $\sim$10Myr of HR8799 would be expected to cause warm dust emission at a level similar to that currently observed, which may provide an explanation for such bright emission in other systems at $\sim10$Myr ages.
	\end{abstract}
	
	
	
	
	\section{Introduction}
	\label{sec:intro}
	The vast number of exo-planets that have been observed in recent years has revolutionised planet formation and evolution theories. However the vast majority of these planets have been detected using transit or radial velocity techniques, which are only sensitive to planets within a few au (at best) of the host star \citep[e.g.][]{2003Msngr.114...20M,2011ApJS..197....8L,2011arXiv1109.2497M,2013A&A...551A..90M,2014ApJ...784...44L,2016ApJ...822...86M,2016ApJ...817..104R,2016ApJ...819...28W}. Currently direct imaging offers the best option for detecting planets outside this limiting distance. However, due to the complexities in nulling the stellar halo with extreme precision, most direct imaging studies are only sensitive to planets above a few Jupiter masses \citep[e.g.][]{2016A&A...587A..55V, 2017A&A...605L...9C}. Nevertheless, the detection of wide orbit giant planets can place stringent constraints on the architecture of the inner planets \citep[e.g.][]{2016MNRAS.455.2980B,2016MNRAS.457..465R,2016arXiv160908058M,2017MNRAS.467.1531H,2017AJ....153...42L,2017MNRAS.469..171R}, and the existence of unseen planets invoked to explain structure observed in debris belts (e.g. $\beta$ Pic \citealt{2009A&A...506..927L}).  
	\begin{table*}
		{
			\centering
			\begin{tabular}{ c c c c c c c c }
				\hline
				Planet & $M$ ($\mathrm{M_J}$) & $a$ (au) & $e$ & $I$ (deg) & $\Omega$ (deg) & $\varpi$ (deg) & $M_A$ (deg)\\
				\hline
				e & 9 $\pm$ 2 & 15.4 $\pm$ 0.2 & 0.13 $\pm$ 0.03 & 25 $\pm$ 3 & 64 $\pm$ 3 & 176 $\pm$ 6 & 326 $\pm$ 5\\ 
				d & 9 $\pm$ 3 & 25.4 $\pm$ 0.3 & 0.12 $\pm$ 0.02 & & & 91 $\pm$ 3 & 58 $\pm$ 3\\  
				c & 9 $\pm$ 3 & 39.4 $\pm$ 0.3 & 0.05 $\pm$ 0.02 & & & 151 $\pm$ 6 & 148 $\pm$ 6 \\
				b & 7 $\pm$ 2 & 69.1 $\pm$ 0.2 & 0.020 $\pm$ 0.003 & & & 95 $\pm$ 10 & 321 $\pm$ 10\\
				\hline
			\end{tabular}
			\captionsetup{justification=centering}
			\caption{Masses and stellocentric orbital elements of the planets around HR8799 from \protect\citealt{2014MNRAS.440.3140G}. The orbital elements refer to semi-major axis, eccentricity, inclination, longitude of ascending node, longitude of pericentre and mean anomaly respectively. The planets were predicted to be coplanar with $I=25\pm3^\circ$ and $\Omega=64\pm3^\circ$.}
			\label{tab:4pl}
		}
	\end{table*}
	
	HR8799 is one of the most well known directly imaged systems, which has been observed to host four giant planets that are a few times the mass of Jupiter \citep{2008Sci...322.1348M,2010Natur.468.1080M}. Debris belts, both internal and external to the planets have also been detected \citep[e.g.][]{2006ApJS..166..351C,2007ApJ...660.1556R,2009ApJ...705..314S,2011ApJ...740...38H,2011A&A...531L..17P,2014ApJ...780...97M,2016MNRAS.460L..10B}. As such, HR8799 is an ideal test system for planet formation and evolution theories, as interactions between planets themselves and with debris can be investigated. Moreover, HR8799 is perhaps currently one of the most promising systems for understanding the formation of the Solar System, as both harbour four giant planets in addition to internal and external populations of debris.      
	
	The outer belt around HR8799 has recently been subject to observations with the \textit{Atacama Large Millimeter Array} (ALMA) \citep{2016MNRAS.460L..10B}. These observations provide the highest resolution images of the outer belt to date and suggest that the position of the inner edge of the belt is too far out to be carved by dynamical interactions with the outermost known planet. \citealt{2016MNRAS.460L..10B} postulate therefore that an additional planet might be present around HR8799, external to the outermost known planet and below current detection thresholds, which is responsible for the structure of the outer planetesimal belt. In this paper we therefore investigate whether a fifth planet around HR8799 indeed provides a better explanation for the ALMA observations of the outer disk, compared with the four known planets in their currently observed configuration. 
	
	In \S\ref{sec:HR8799} we give specific details of the HR8799 system including the planets and debris belts. In \S\ref{sec:modeling4pl} we use N-body simulations to model how the four known planets interact with the outer planetesimal belt and how well this predicts the ALMA observations. In \S\ref{sec:fifthpl} we include an additional planet in our simulations, which is external to the outermost known planet, to investigate whether the predicted outer belt agrees more strongly with the ALMA observations. In \S\ref{sec:inwarddelv} we consider whether this additional planet can replenish the mass of the inner belt, through inward scattering of material from the outer planetesimal belt, before finally summarizing and concluding in \S\ref{sec:conc}.

	\section{HR8799}
	\label{sec:HR8799}
	\subsection{Stellar properties}
	HR8799 is an A type star \citep{2003AJ....126.2048G} at a distance of 39.4pc \citep{2007A&A...474..653V}. Characteristic pulsations in luminosity and an unusual deficiency in iron peak elements compared with similar type stars place it as a $\gamma$-Doradus and $\lambda$-Bootis type star respectively \citep{1990ApJ...363..234V,1999AJ....118.2993G,1999PASP..111..840K,1999MNRAS.309L..19H}. The mass, radius and luminosity estimates of HR8799 most commonly referenced in the literature are $\sim$1.5$\mathrm{M_\odot}$, 1.44$\mathrm{R_\odot}$ and 5.05$\mathrm{L_\odot}$ respectively \citep{2008Sci...322.1348M,2010Natur.468.1080M,2012ApJ...761...57B}. While consensus has largely been reached on these fundamental stellar parameters, the age of the system, especially since the discovery of the giant planets, remains a topic of much debate. The age of HR8799 is of vital importance in determining the mass and therefore the nature of the planets, as planetary evolution models predict planets to cool and therefore significantly dim over time. A younger age estimate for HR8799 therefore predicts less massive planets and vice versa. Most studies agree that HR8799 is unequivocally young at $<100$Myr, however estimates of $\sim$Gyrs do exist \citep{2010MNRAS.405L..81M}. A summary of age estimates from a variety of different techniques can be found in Table 1. of \citealt{2012ApJ...761...57B}. 
	
	In the discovery papers of the directly imaged planets \citep{2008Sci...322.1348M,2010Natur.468.1080M}, the age of HR8799 was assumed to be $60^{+100}_{-30}$Myr due to: 1) the galactic space motion of HR8799 placing it as a likely member of the Columba moving group which contains stars with ages between $\sim$30-40Myr \citep{2008hsf2.book..757T, 2011ApJ...732...61Z,2016IAUS..314...41B}. 2) The position of HR8799 on a Hertzsprung-Russell diagram is similar to stars with ages of $\sim50-70$Myr \citep{2001ARA&A..39..549Z}. 3) The fact that $\gamma$-Doradus and $\lambda$-Bootis type stars are typically young with ages of $\sim$100Myr \citep{1995MNRAS.277.1404K,2002AJ....124..989G}. 4) The probability of detecting a debris disk decreases with age, suggesting that HR8799 is indeed young \citep{2001ApJ...555..932S,2003ApJ...598..636D,2005ApJ...620.1010R,2007ApJ...660.1556R}. We note that both support and doubt of this reasoning has been cast by a variety of authors \citep[e.g.][]{2010MNRAS.405L..81M,2010MNRAS.406..566M,2010ApJ...721L.199M,2010ApJ...716..417H,2011ApJ...732...61Z,2011ApJ...729..128C,2012ApJ...761...57B}. Notably, support for the young age of HR8799 is given by measurements of the luminosity and radius of the star using the CHARA Array Interferometer, which places the age at $33^{+7.0}_{-13.2}$Myr, assuming that HR8799 is contracting toward the zero-age main sequence \citep{2012ApJ...761...57B}. Moreover, the probability that HR8799 is a member of the Columba association using the Banyan II online tool detailed in \citealt{2014ApJ...783..121G} is 75\%, assuming weighted priors and that the star is younger than 1Gyr. Studies have cast some doubt on the Columba membership of HR8799 however, notably by \citealt{2010ApJ...716..417H} who suggest that HR8799 is too far from the centroid of Columba to be a likely member. However Columba association members are mainly southern hemisphere targets, such that a northern hemisphere target like HR8799 is likely to be significantly separated from these objects. Indeed, since \citealt{2010ApJ...716..417H}, more members of the Columba association have been identified, with relative positions closer to HR8799 \citep[e.g.][]{2011ApJ...732...61Z,2013ApJ...774..101R}.

	\subsection{Planets}
	For an assumed age of $60^{+100}_{-30}$Myr, the masses of the four planets around HR8799 from the discovery papers were predicted to be $7^{+4}_{-2}$, $10\pm3$, $10\pm3$, $10\pm3$ $\mathrm{M_J}$ for planets b, c, d, e respectively \citep{2008Sci...322.1348M,2010Natur.468.1080M}, where b is the outermost planet and e the innermost. That is, the planets are indeed planetary objects rather than brown dwarfs. As these planets have such large masses, many authors have investigated the dynamical stability of such a planet configuration \citep[e.g.][]{2009MNRAS.397L..16G,2009A&A...503..247R,2010ApJ...710.1408F,2011ApJ...741...55S,2012ApJ...755L..34C,2012ApJ...755...38S,2013A&A...549A..52E,2014MNRAS.440.3140G}. Indeed, the dynamical stability of the planets can place an additional constraint on the age of the system. Many independent studies agree that stability is maintained between the planets for the lifetime of the system likely due to the planets being in a 1b:2c:4d:8e mean motion resonant chain (e.g. \citealt{2009MNRAS.397L..16G,2009A&A...503..247R,2010ApJ...710.1408F,2012ApJ...755...38S,2014MNRAS.440.3140G,2016AJ....152...28K,2016A&A...587A..57Z}, see \citealt{2015ApJ...803...31P,2016A&A...592A.147G} for studies which show that stability can be maintained without resonances however). That is, for every 1 orbit of b, c orbits twice etc. Such a configuration is supported by recent work from \citealt{2014MNRAS.440.3140G} who consider a suite of masses and orbital elements for the planets to see which simultaneously remain stable and best reproduce the astrometric observations of the planets. We show the masses and stellocentric orbital elements for each of the planets from the best fitting model from \citealt{2014MNRAS.440.3140G} in Table \ref{tab:4pl} (equivalent to Table 1. in \citealt{2014MNRAS.440.3140G}). This model predicted the planets to be coplanar with an inclination and longitude of ascending node of $I=25\pm3^\circ$ and $64\pm3^\circ$ respectively. We note however that it is unclear whether the planets are indeed coplanar, as two independent studies find that at least planet d may not be co-planar with the other planets (\citealt{2012ApJ...755L..34C, 2015ApJ...803...31P}). Co-planarity of the planets is supported by recent fitting of astrometric data detailed in \citealt{2016AJ....152...28K}, however here some astrometry points considered by earlier works are omitted.

	\subsection{Disk Structure}
	\label{subsec:diskstruc}
	Belts of material have also been observed around HR8799, both internal and external to the known planets \citep[e.g.][]{2006ApJS..166..351C,2007ApJ...660.1556R,2009ApJ...705..314S,2011ApJ...740...38H,2011A&A...531L..17P,2014ApJ...780...97M,2016MNRAS.460L..10B}. $Spitzer$ and $Herschel$ observations find an unresolved belt of warm ($\sim$150K) dust internal to the known planets between $\sim6-15$au, with a mass of $1.1\times10^{-6}\mathrm{M_\oplus}$ in small grains between 1.5-4.5$\mathrm{\mu m}$ \citep{2009ApJ...705..314S,2014ApJ...780...97M}. External to the known planets far-infrared and mm observations resolve a cold ($\sim$45K) planetesimal belt between $\sim100-300$au \citep[e.g.][]{2009ApJ...705..314S,2011ApJ...740...38H,2011A&A...531L..17P,2014ApJ...780...97M} with a mass in 10-1000$\mathrm{\mu m}$ dust grains of $1.2\times10^{-1}\mathrm{M_\oplus}$ \citep{2009ApJ...705..314S}. Further outside the planetesimal belt, a halo of small grains has also been observed to extend out to $\sim1000$au \citep[e.g.][]{2009ApJ...705..314S,2014ApJ...780...97M}. 
	
	Recent ALMA observations of the outer planetesimal belt resolve it to be between $\sim$145-429au. We defer the reader to \citealt{2016MNRAS.460L..10B} for a detailed description of the ALMA observations and modelling of this outer belt, however we highlight the main points here. Figure 1 in \citealt{2016MNRAS.460L..10B} shows the dirty image (referring to the inverse Fourier transform of the visibilities) of the continuum emission of HR8799 at 1.34mm and the dirty beam, which had a beam size of 1.7 $\times$ 1.3 arcsec$^2$ in RA and DEC respectively for a position angle of $89^\circ$ (taken anti-clockwise from North). From this dirty image, it is clear that there is a ring of emission at $\sim$5.5arcsec from the star, indicating the presence of an outer planetesimal disk around HR8799.
	
	In order to estimate parameters of this outer disk, \citealt{2016MNRAS.460L..10B} modelled the disk emission in the image space using 6 parameters: (1) the radius of the inner ($R_\mathrm{in}$) and (2) outer ($R_\mathrm{out}$) edges, (3) a value $\Gamma$ which defines how the optical depth varies as $r^\Gamma$, (4) the inclination of the disk from face on ($I_\mathrm{disk}$), (5) a position angle measured anti-clockwise from North ($\theta$) and (6) the total flux density of the disk at 1.34mm.  
	
\begin{figure}
		\centering
		\includegraphics[trim={0.0cm 0cm 0cm 0cm},width=0.5\textwidth]{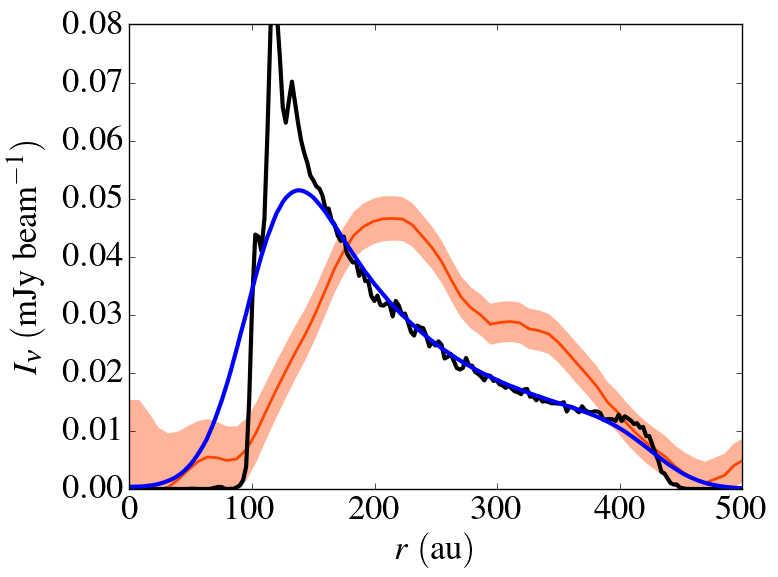}
		\caption{Intensity profile of outer disk around HR8799 from ALMA observations, shown by the orange line with the shaded region referring to the $1\sigma$ rms of the noise per beam. The black line gives the intensity profile of the outer disk predicted by our simulations from the four known planets interacting with a population of test particles. The blue line shows this profile once the simulated image of the outer belt has been convolved with the beam of ALMA. It is clear that the profile predicted by the four planets in isolation is not a good fit to the ALMA observed profile.}
		\label{fig:4plintens}
	\end{figure}

	From running a MCMC minimisation procedure, the 6 parameters which produced a model disk which most closely represented the dirty image of HR8799 are given in Table 2 in \citealt{2016MNRAS.460L..10B}, with $R_\mathrm{in}=145^{+12}_{-12}$au, $R_\mathrm{out}=429^{+37}_{-32}$au, $\Gamma=-1.0^{+0.4}_{-0.4}$, $I_\mathrm{disk}=40^{+5\circ}_{-6}$ and $\theta=51^{+8\circ}_{-8}$. The azimuthally averaged intensity profile of a model disk with these best fitting parameters in units of mJy beam$^{-1}$, is shown by the orange line in Figure \ref{fig:4plintens} (Booth priv comm). Herein we simply refer to this intensity profile as the observed intensity profile for the outer planetesimal belt around HR8799, noting that this profile is actually the profile of the best fitting model image from \citealt{2016MNRAS.460L..10B} convolved with the ALMA beam with the residuals added on top, rather than from the dirty image of HR8799 itself. The shaded regions around the intensity profile refer to the 1$\sigma$ rms of the noise per beam at a given radial location (extracted using a method described in \citealt{2016MNRAS.460.2933M}).
	
	Notably, \citealt{2016MNRAS.460L..10B} suggest the position of the inner edge of the outer belt cannot be explained by dynamical interactions with planet b in its current configuration and therefore, this might be indicative of an additional, yet to be detected planet which is external to planet b.
	
	\begin{figure}
		\centering
		\includegraphics[width=0.5\textwidth]{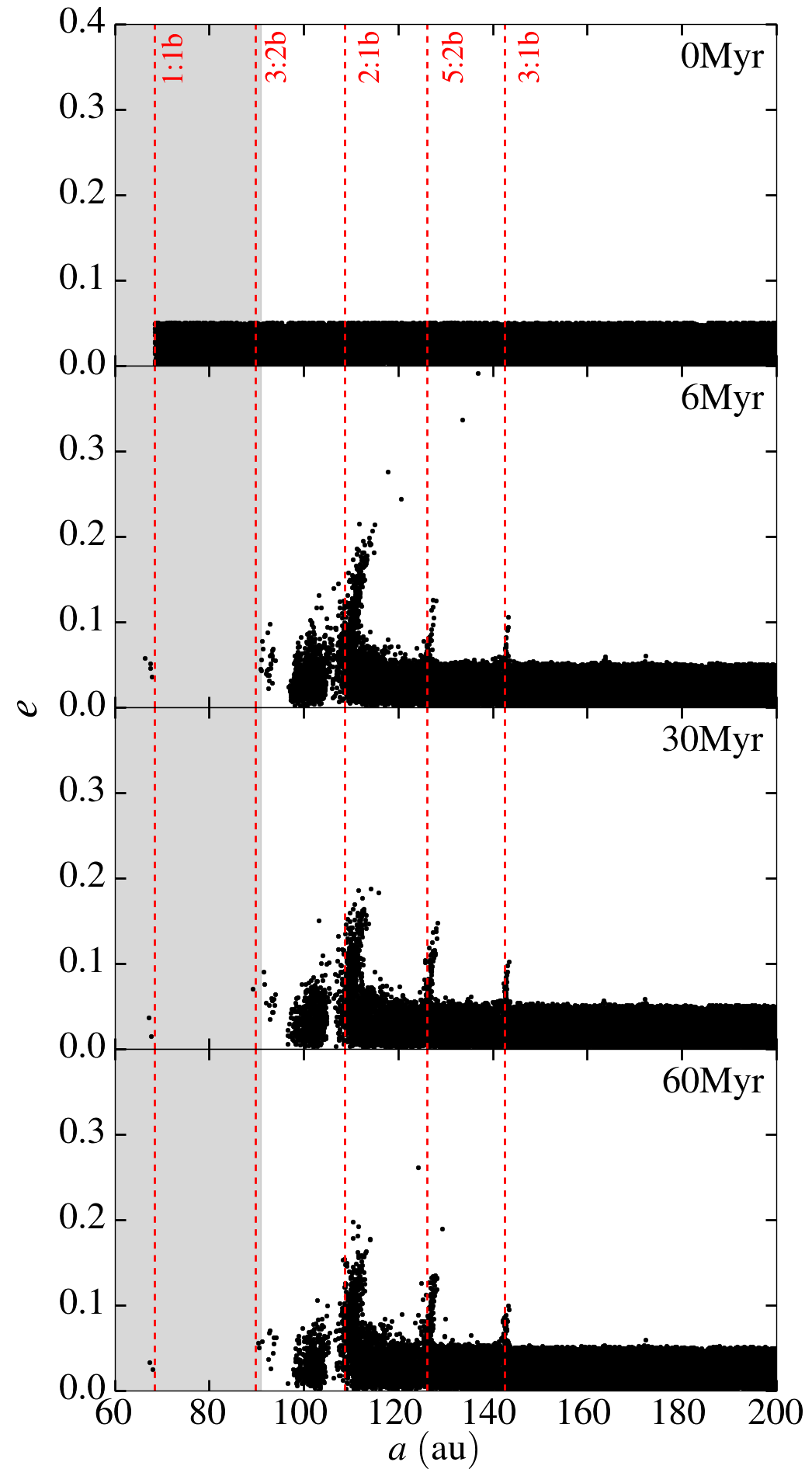}
		\caption{Semi-major axis vs. eccentricity of particles evolving due to dynamical interactions with the four known planets around HR8799. The grey shaded area refers to the classical chaotic zone around HR8799b. The red dashed lines refer to the mean motion resonances of HR8799b.}
		\label{fig:evsa}
	\end{figure}
	\section{Outer disk interaction with the four known planets}
	\label{sec:modeling4pl}
	\subsection{Simulations}
	\label{subsec:modelling}
	We first model how an outer planetesimal belt around HR8799 would interact with the four known planets. We assume masses and stellocentric orbital elements of the planets equal to those derived in \citealt{2014MNRAS.440.3140G}, as shown in Table \ref{tab:4pl}. We also assume the same stellar mass and radius of HR8799 used in \citealt{2014MNRAS.440.3140G} of $1.56\mathrm{M_\odot}$ and $1.44\mathrm{R_\odot}$ respectively \citep{2010Natur.468.1080M,2012ApJ...761...57B}. We represent the outer disk as a population of 50,000 non-interacting massless particles which are initially distributed evenly between the semi-major axis of HR8799b (69.1au) and the outer edge of the disk derived from ALMA observations (429au \citealt{2016MNRAS.460L..10B}). The particles are initialised with respect to the centre of mass of the system and the orbital plane of the planets, with eccentricities and inclinations randomly sampled between 0-0.05 and 0-0.05radians respectively. The remaining angular orbital elements are initialised randomly between 0-2$\pi$. For the radii of the four known planets we use the 'Jovian Worlds' mass to radius relation from \citealt{2017ApJ...834...17C}, valid for 0.414$\mathrm{M_J}$ $<$ $M$ $<$ 0.08$\mathrm{M_\odot}$\footnote{\label{note1}We use the deterministic version of the probabilistic mass to radius relation from \citealt{2017ApJ...834...17C}.}: 
	\begin{equation}
	\frac{R}{R_\oplus}=17.74\left(\frac{M}{M_\oplus}\right)^{-0.044}.
	\label{eq:radtomass1}
	\end{equation}
	This gives a radii of 1.11$\mathrm{R_J}$ for c, d, e and 1.12$\mathrm{R_J}$ for b respectively (we note however that the choice of planet radii has little effect on the results of this work). 
	
	To model the gravitational interaction between all bodies, we use the N-body integrator REBOUND \citep{2012A&A...537A.128R}, using a hybrid integrator (HERMES) which switches from a fixed, to a variable time-step integrator \citep{2015MNRAS.446.1424R,2015MNRAS.452..376R} when a pair of objects orbit within a given distance. This allows for computational efficiency when bodies are widely spaced and for close encounters to be followed accurately. For the fixed time-step integrator we use a time-step of 2.4yrs (5\% of the period of HR8799e) and switch to the variable time-step integrator when particles orbit within 8 Hill radii of one of the planets or within 50 stellar radii of the star. We integrate the system for 60Myr, noting the orbital elements of all bodies at intervals of 0.6Myr. Particles are removed if they are ejected from the system, which is defined as reaching a distance of 10,000au from the center of mass, or have a physical collision with any of the planets or the star. During a collision, the time and the orbital elements of the particle are noted.
	
	\subsection{Results}
	\label{subsec:4plresult}
	The eccentricity of particles against semi-major axis out to 200au at four different epochs is shown in Figure \ref{fig:evsa}. Close to HR8799b, overlapping first-order resonances induce chaotic motions in particles and they are quickly removed from the simulation. The chaotic zone around a planet extends to an approximate distance of $\Delta a=1.5a\mu^{2/7}$
	where $a$ is the semi-major axis of the planet and $\mu$ is the planet to star mass ratio, $M/M_\star$, where $M$ and $M_\star$ are the planet and star mass respectively \citep[e.g.][]{1980AJ.....85.1122W,1989Icar...82..402D}. For HR8799b, the width of the chaotic zone extends out to $\sim90$au, which is shown by the grey shaded region in Figure \ref{fig:evsa}. Particles are also perturbed at the mean motion resonances (MMR) of HR8799b (red dashed lines), with the positions of these resonances given by $a = a_\mathrm{b}(p/q)^{2/3}$, where $p$ and $q$ are integers, $p>q$ and $a_\mathrm{b}$ is the semi-major axis of HR8799b. We note that the same evolution is also seen in the inclinations of the particles, with the maximum inclination after 60Myr for non-scattered particles reaching $\sim$10$^\circ$ at the 2:1b MMR. Beyond 150au, we find particles do not significantly evolve over 60Myr. Secular interactions at these distances are not significant, with particles being perturbed by negligibly small forced eccentricities and inclinations.
	
	To investigate the presence of any smaller scale resonant structure in the surface density of the particles, we plot the de-projected $x$ and $y$ positions of particles in the top panel of Figure \ref{fig:stacked4pl}. As few particles undergo scattering interactions with the planets after $\sim$30Myr (see Figure \ref{fig:evsa}), we take output from the last 16 intervals of the simulation (from 51-60Myr) and combine them, by choosing a frame co-rotating with HR8799b. This substantially boosts the signal to noise of the image, as the number of particles considered increases from 5$\times$10$^4$ to 8$\times$10$^5$. The white cross gives the position of HR8799b for reference. This image highlights that no significant structure is seen external to the inner edge, other than a small dip caused by the 2:1b MMR.

	\subsection{Comparing outer disk with ALMA observations}
	\label{subsec:4plalmacomp}
	\begin{figure}
		\centering
		\includegraphics[trim={0.0cm 0cm 0cm 0cm},width=0.5\textwidth]{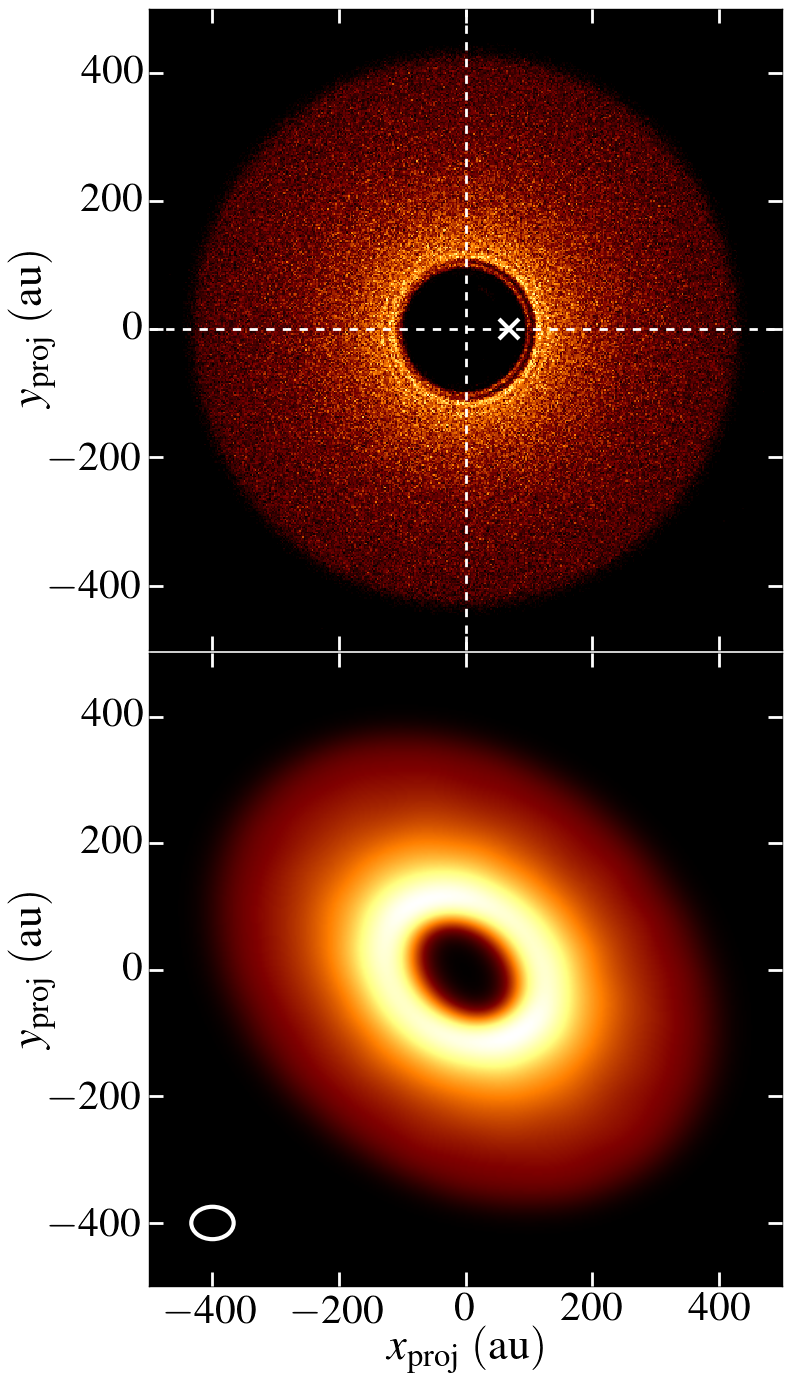}
		\caption{(\textit{top}) Surface density image of particles after 60Myr due to interactions with the four known planets. Particles from 16 epochs between 51-60Myr are stacked together in a frame co-rotating with HR8799b (white cross). (\textit{bottom}) Top panel scaled from surface density to intensity, which is inclined by and given a position angle of 40$^\circ$ and 51$^\circ$ respectively, which is then convolved with the beam size of ALMA (1.7 $\times$ 1.3 arcsec$^2$, white ellipse, \protect\citealt{2016MNRAS.460L..10B}).}
		\label{fig:stacked4pl}
	\end{figure}
	It is clear therefore from Figure \ref{fig:evsa} that the four known planets carve an inner edge of the planetesimal belt at $\sim$100au. However, as discussed in \S\ref{subsec:diskstruc}, ALMA observations of HR8799 suggest that the inner edge is much further out at 145au (\citealt{2016MNRAS.460L..10B}, Figure \ref{fig:4plintens}). To see how well our simulations predict the overall structure of the disk observed by ALMA, rather than just the inner edge, we compare our simulations directly with the ALMA observations. 
	
	To compare our simulated surface density image of HR8799 (e.g. top panel of Figure \ref{fig:stacked4pl}) with the ALMA data, we must convert it to an intensity image, with the disk inclined from face on by $I=40^\circ$ and a position angle (anti-clockwise from North) of $\theta=51^\circ$ (\S\ref{subsec:diskstruc}, \citealt{2016MNRAS.460L..10B}). We assume that the intensity of the disk is given by:
	\[
	I_\nu(r) \propto \Sigma(r) B_\nu(T)\kappa_\nu,
	\]
	where $\Sigma (r)$ is the surface density of the disk at a radial position $r$, $B_\nu(T)$ is the Planck function at temperature $T$ and $\kappa_\nu$ is the opacity. We assume that the opacity is a fixed quantity and that planetesimals that are emitting at 1.3mm absorb and emit like blackbodies ($T \propto r^{-1/2}$), allowing the Planck function to be approximated in the Rayleigh Jeans limit, resulting in 
	\begin{equation}
	I_\nu = K\Sigma(r) r^{-1/2},
	\label{eq:Kscal}
	\end{equation} 
	where $K$ is a scaling factor. We therefore take the simulated image of HR8799 from the top panel of Figure \ref{fig:stacked4pl}, scale it by a factor of $r^{-1/2}$ to convert it to intensity, incline it (from face on) by $I=40^\circ$ and rotate it to have a position angle (anti-clockwise from North) of $\theta=51^\circ$.
	
	For the simulated intensity image to have the same resolution expected from ALMA, it is necessary to convolve it with the beam of ALMA. We assume that the beam can be approximated by an elliptical Gaussian with a FWHM in $x$ and $y$ of 1.7 and 1.3arcsec respectively (equal to size of the beam in RA and DEC respectively from \citealt{2016MNRAS.460L..10B}), and to 67.0 and 51.2au respectively at the distance of HR8799. The bottom panel of Figure \ref{fig:stacked4pl} shows the simulated surface density image from the top panel once it has been scaled to intensity, inclined, rotated and convolved with the beam of ALMA. The beam is shown by the white ellipse for reference. An azimuthally averaged radial profile of the bottom panel of Figure \ref{fig:stacked4pl}, which is calculated by using a series of commonly aligned elliptical apertures with an equivalent inclination and position angle of $I=40^\circ$ and $\theta=51^\circ$ respectively, therefore gives an intensity profile which can be compared with the ALMA observed profile (orange line in Figure \ref{fig:4plintens}). 
	
	We also consider different initial radial profiles for the surface density of the particles. Thus far, the particles were initially uniformly distributed in semi-major axis, equivalent to an initial surface density proportional to $r^{-1}$. We consider whether changing this initial distribution of particles improves the agreement between the simulated and observed intensity profiles. To do this, we weight the contribution of a particle in the simulated intensity image by $a_{\mathrm{ini}}^{-\gamma}$, where $a_\mathrm{ini}$ is the initial semi-major axis of the particle and $\gamma$ is a scaling factor. A zero value of $\gamma$ would therefore be equivalent to the particles being initially uniformly distributed in semi-major axis, giving an initial surface density proportional to $r^{-1}$.
	
	We subsequently conduct a simultaneous $\chi^2$ minimisation over a range of values of $\gamma$ and the vertical scaling factor ($K$ from eq. (\ref{eq:Kscal})) to find the combination of these two parameters which causes the strongest agreement between the simulated intensity profile and the one observed by ALMA. Each $\chi^2$ value in this minimisation takes the standard value of 
	\[\chi^2=\frac{N_\mathrm{ind}}{N}\sum\limits_{1}^N{\frac{(O(r)-M(r))^2}{\sigma(r)^2}},\]
	where $O(r)$ and $M(r)$ are the values of the observed and simulated intensity profiles respectively at a radial position $r$, $\sigma(r)$ is the 1$\sigma$ rms of the noise per beam at a given radial location for the observed profile (orange shaded region, Figure \ref{fig:4plintens}), $N_\mathrm{ind}\sim7$ is the number of ALMA beams that fit within our image size and $N=64$ is the number of points in the observed profile (orange line Figure \ref{fig:4plintens}). This minimisation gives $\gamma=0$, i.e. that the best fitting simulated intensity profile is one where particles are initially uniformly distributed in semi-major axis. The simulated intensity profile for $\gamma=0$ (with the associated best fitting value of $K$), both with and without convolving our simulated intensity image with the beam of ALMA are given by the blue and black curves respectively in Figure \ref{fig:4plintens}.
	
	It is clear that the intensity profile from the simulation does not fit well with that observed by ALMA. This therefore supports the conclusion in \citealt{2016MNRAS.460L..10B} that the four planets in their current configuration cannot shape an outer disk consistent with the one observed by ALMA, in particular the position of the inner edge.

	\section{Including an additional fifth planet}
	\label{sec:fifthpl}
	\subsection{Stability of an additional planet}
	\label{subsec:stability}
	\begin{figure}
		\centering
		\includegraphics[trim={0.0cm 0cm 0cm 0cm},width=0.5\textwidth]{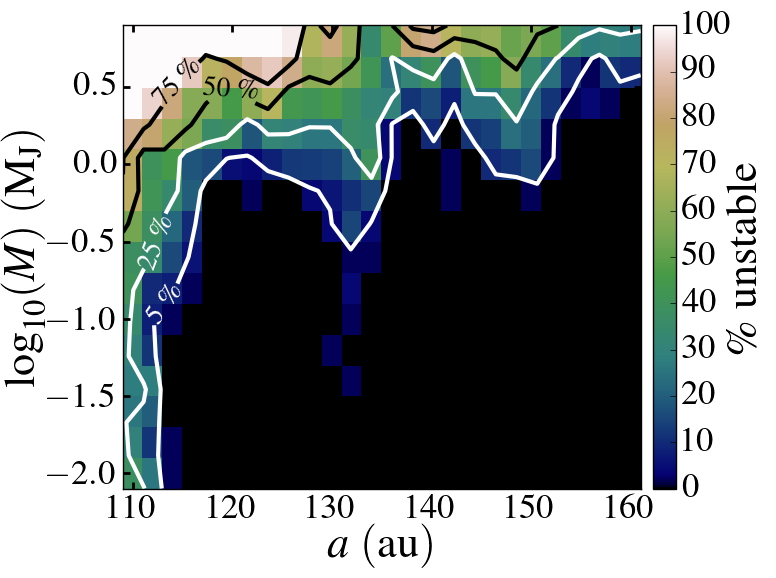}
		\caption{Percentage of 50 simulations that went unstable when including an additional fifth planet with a given mass and semi-major axis to HR8799. The angular orbital elements for a given fifth planet were randomly sampled for each of the 50 simulations. Contours refer to specific percentages.}
		\label{fig:stabil}
	\end{figure}
	We therefore consider whether a fifth undetected planet around HR8799, external to the known planets, predicts an outer disk with an intensity profile that more closely represents the one observed by ALMA. We consider a fifth planet with masses of 0.01, 0.016, 0.025, 0.04, 0.063, 0.1, 0.16, 0.25, 0.4, 0.63, 1.0, 1.6, 2.5, 4.0, 6.3$\mathrm{M_J}$ (linear in log space) and semi-major axes ranging from 110 to 160au with a spacing of 2au. This planet is initialised on a circular orbit with respect to the star (for consistency with the initial orbital elements of the known planets, Table \ref{tab:4pl}) and is coplanar with the orbits of the 4 known planets. 
	
	Due to the large masses of the known planets around HR8799, one must consider how stable an additional planet would be if it were present. Indeed the known planets are likely stable due to the complex resonant structure of their orbits, which might be disrupted by an additional planet. For each mass and semi-major axis of the fifth planet, we run 50 simulations where the initial true anomaly and longitude of pericentre of the fifth planet are randomized between 0-2$\pi$. We note that randomizing the longitude of pericentre is necessary to account for the small initial eccentricity of the fifth planet with respect to the barycentre of the system. The parameters of the simulation are identical to that described in \S\ref{subsec:modelling}, however for computational efficiency we do not include the population of massless particles. 
	
	The percentage of the 50 simulations which went unstable for each fifth planet is shown in Figure \ref{fig:stabil}, with contours highlighting specific percentages. Systems are defined to be unstable if the semi-major axis of any of the planets exceeds 10\% of its initial value during the simulation. We find that simulations go unstable due to multiple planets being scattered from their original orbits rather than just a single planet. This is perhaps expected, as the known planets orbit on the edge of stability in a resonant chain, such that if this chain is disrupted by the scattering of a single planet, it will likely affect all the other planets in the resonant chain as well. As might also be expected, Figure \ref{fig:stabil} shows that a more massive planet closer to HR8799b results in a system that is more likely to go unstable. Moreover a fifth planet is more likely to cause an instability in the system if it approaches the 2:1b, 5:2b, 3:1b MMRs of HR8799b at $\sim$110, 130 and 145au respectively. 
	
	We note that due to the resolution of our grid, the fifth planets we consider do not exactly lie on the major MMRs of HR8799b. We therefore we do not also sample whether a fifth planet would be stable (in addition to those shown in Figure \ref{fig:stabil}) as part of a resonant chain with the known planets. Indeed it is not unreasonable to think that the process which caused the known planets to be caught in a resonant chain might also extend to an additional planet. We leave the topic of whether expanding the resonant chain of the known planets produces stability zones for additional planets to future work.
	\begin{figure*}
		\centering
		\includegraphics[trim={0.0cm 0cm 0cm 0cm},width=1\textwidth]{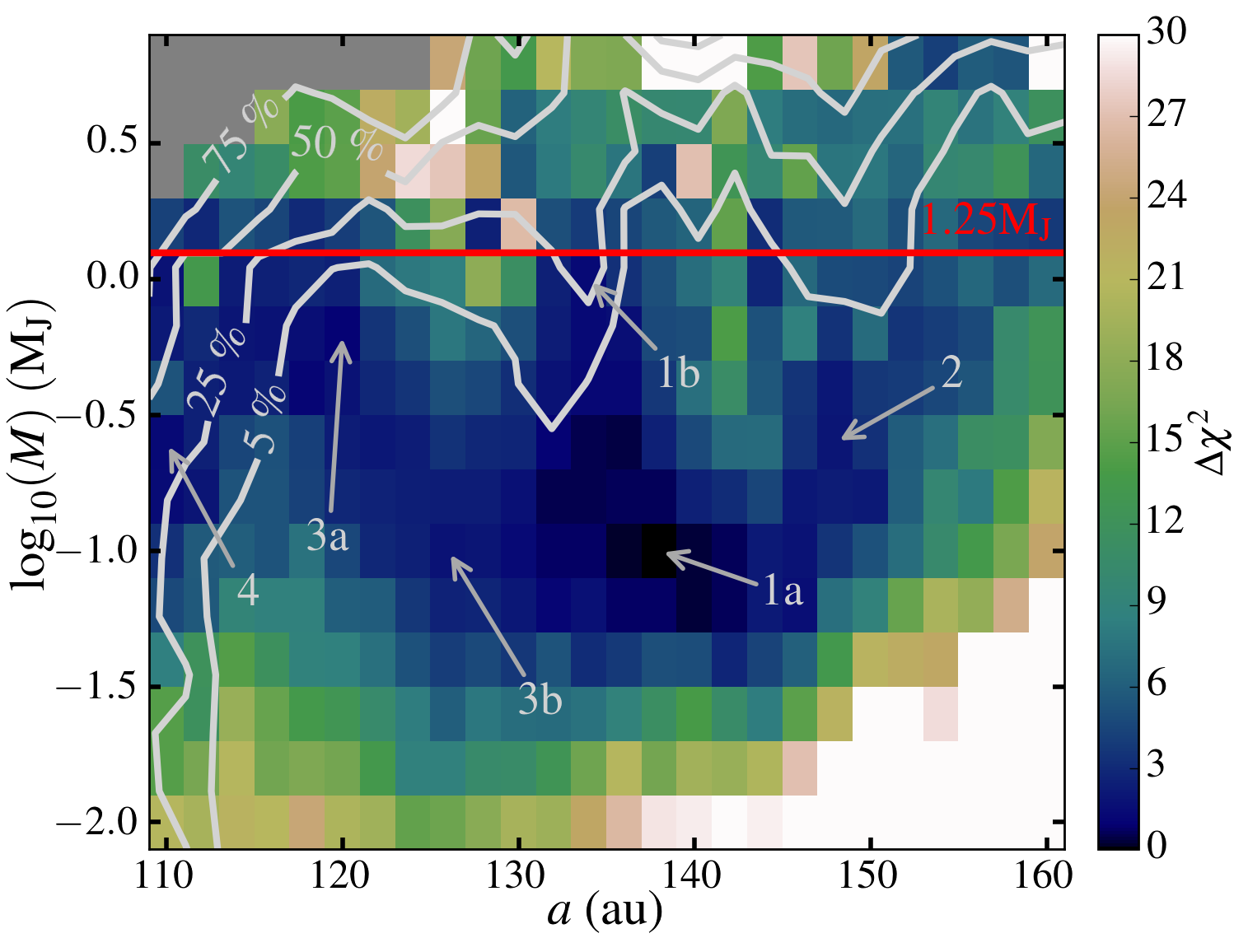}
		\caption{How well intensity profiles of the outer disk of HR8799 from our simulations with an additional fifth planet with given mass and semi-major axis agree with the observed intensity profile from ALMA. The colour scale gives the $\chi^2$ value of the fit between the simulated and observed profiles relative to the $\chi^2$ value of the profile that best fit the observed profile. The light grey contours are the stability contours shown in Figure \ref{fig:stabil}, with the red line giving the direct imaging threshold for planet detection. The greyed out region refers to fifth planets where 100\% of the simulations from Figure \ref{fig:stabil} were unstable. Intensity profiles of individual annotations are shown in Figure \ref{fig:gamma_curves}.}
		\label{fig:chi2}
	\end{figure*}
	\subsection{Simulations}
	We now consider the effect each of the stable fifth planets has on the population of particles initially described in \S\ref{subsec:modelling}. Fifth planets close to HR8799b, with a high enough mass for all of the 50 stability simulations in Figure \ref{fig:stabil} to go unstable were disregarded from further study. For each fifth planet we select a true anomaly and longitude of pericentre that produced a stable configuration from Figure \ref{fig:stabil} and initialise it with a population of test particles as described in \S\ref{subsec:modelling}. The radii of each of the fifth planets are given by the mass to radius relation from eq. (\ref{eq:radtomass1}) for 0.414$\mathrm{M_J}$ $<$ $M$ $<$ 0.08$\mathrm{M_\odot}$. For 2.04$\mathrm{M_\oplus}$ $<$ $M$ $<$ 0.414$\mathrm{M_J}$, we use the 'Neptunian Worlds' relation from \citealt{2017ApJ...834...17C}$^{\ref{note1}}$:
	\begin{equation}
	\frac{R}{R_\oplus}=0.81\left(\frac{M}{M_\oplus}\right)^{0.589}.
	\label{eq:radtomass2}
	\end{equation}
	
	For each fifth planet we followed the same prescription discussed in \S\ref{subsec:modelling}, running simulations to compare the intensity profile of the disk from our simulations with the profile observed by ALMA. For computational efficiency however, we first considered whether the number of particles could be reduced from 50,000 without degrading the quality of a given intensity profile at the end of our simulations. We took 9 of the considered fifth planets, which sampled the overall range of masses and semi-major axes and ran these simulations to 60Myr with 50,000 particles. We then artificially removed particles to reduce the quality of the intensity profile. We found that 10,000 particles could be used without significantly degrading the final intensity profile for the 9 considered fifth planets. For the remainder of the simulations for all the fifth planet masses and semi-major axes described in \S\ref{subsec:stability} we therefore include 10,000 particles. 
	
	To off-set any reduction in the signal to noise of the simulated images generated by each fifth planet from reducing the number of particles, at the end of our simulations we sample each particle 1000 times around its orbit. We note that this approximation is only valid if there is no asymmetric disk structure arising from correlated particle true anomalies, as this structure would become smoothed out. We discuss the validity of this approximation in \S\ref{subsec:resultdiscuss}. 
	
	\subsection{Comparing simulations with ALMA observations}
	\label{subsec:getresult}
	After generating an intensity profile for each fifth planet simulation, we perform a simultaneous $\chi^2$ minimization procedure over the scaling factors $\gamma$ and $K$ in the exact same way discussed in \S\ref{subsec:4plalmacomp}. That is, finding the optimal scaling factor, $\gamma$ (which weights particles in the simulated intensity image according to their initial semi-major axis) and the optimal vertical scaling factor, $K$ from eq. (\ref{eq:Kscal}), which cause the intensity profile from the ALMA convolved simulated intensity image to best fit the observed intensity profile. We note the lowest $\chi^2$ value from this minimisation for each fifth planet considered. 
	
	We find that of all our simulations a fifth planet with a mass of 0.1M$_\mathrm{J}$ and a semi-major axis of 138au predicts an intensity profile that most strongly agrees with the one observed by ALMA. Figure \ref{fig:chi2} shows the minimum $\chi^2$ value of all the other fifth planet parameters relative to this overall $\chi^2$ minimum. We refer to this relative $\chi^2$ value as $\Delta\chi^2$. We include the stability contours from Figure \ref{fig:stabil} for reference. We also plot the current upper mass limit for detection of planets around HR8799 from direct imaging, equivalent to $\sim$1.25$\mathrm{M_J}$ (see Figure 2 of \citealt{2015A&A...576A.133M}), with the red line. We note that the upper mass limit for detection presented in \citealt{2015A&A...576A.133M} only goes out to $\sim80$au, whereas the total field of view for the instrument used in their work included separations out to $\sim275$au. Outside of $\sim80$au however, their contrast sensitivity is limited by thermal background and not by stellar speckles, such that the $\sim$1.25$\mathrm{M_J}$ detection limit is roughly constant for separations $>80$au (Maire priv comm). The greyed out region refers to fifth planets which were 100\% unstable from Figure \ref{fig:stabil}.
	
	\begin{figure*}
		\centering
		\includegraphics[trim={0.0cm 0cm 0cm 0cm},width=1\textwidth]{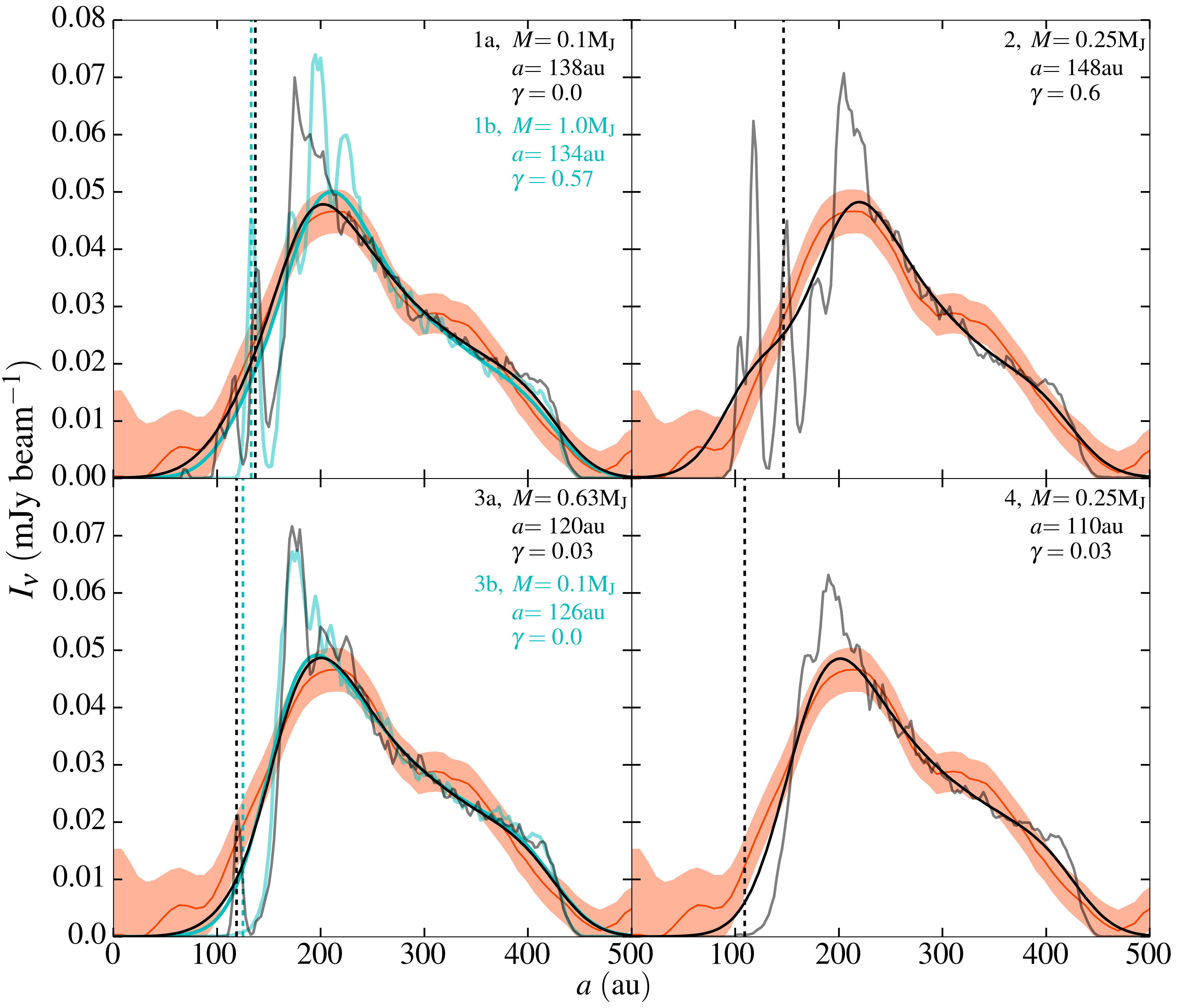}
		\caption{Comparison of intensity profiles generated by selected fifth planets from Figure \ref{fig:chi2} with and without convolving with the beam of ALMA (black and grey lines respectively, also solid and lighter blue lines in the top and bottom left panels), with the profile observed by ALMA (orange line). The orange shaded region gives the 1$\sigma$ rms of the noise per beam. The dashed vertical lines give the position of the fifth planet considered. The values of $\gamma$ refer to the initial scaling of the surface density of particles such that $\Sigma(r) \propto r^{-1}a_\mathrm{ini}^{-\gamma}$. All profiles fit the observed profile significantly better compared with the four known planets in isolation.}
		\label{fig:gamma_curves}
	\end{figure*}
	\subsection{Results}
	\label{subsec:5planresult}
	The minima in the $\Delta\chi^2$ values in Figure \ref{fig:chi2} can be compartmentalised into four main regions, which are annotated by specific example planets discussed below in Figure \ref{fig:chi2}:
	
	1) Fifth planets with a mass and semi-major axis around 0.1$\mathrm{M_J}$ and 138au respectively. The intensity profile of a fifth planet with exactly this mass and semi-major axis (1a in Figure \ref{fig:chi2}) is shown in the top left panel of Figure \ref{fig:gamma_curves}. The grey and black lines refer to the intensity profile before and after the simulated intensity image is convolved with the beam of ALMA respectively. The orange line shows the observed ALMA intensity profile with the shaded region giving the 1$\sigma$ rms of the noise per beam (\S\ref{subsec:4plalmacomp}). The black dashed line shows the semi-major axis of the fifth planet. This fifth planet sculpts an inner edge in the disk at $\sim$160au. However there is also a significant proportion of particles caught in co-rotation with the planet. This results in an intensity profile, which when convolved with the beam of ALMA, is in good agreement with the observed profile. 
	
	2) Fifth planets with semi-major axes between 140-160au, for increasing masses between 0.04-1$\mathrm{M_J}$. The intensity profile for one such fifth planet with a mass and semi-major axis of 0.25$\mathrm{M_J}$ and 148au respectively (2 in Figure \ref{fig:chi2}) is shown in the top right panel of Figure \ref{fig:gamma_curves}. The lines refer to the same quantities as the top left panel. This planet carves a gap with an outer edge at $\sim200$au, with a significant number of particles surviving internal to the orbit of the fifth planet. A population of particles is also present in co-rotation with the planet. Once the simulated intensity image is convolved with the beam of ALMA, these different particle populations produce a profile that has an overall good agreement with that observed by ALMA. However the inner edge is not fit as well as fifth planets from region (1). 
	
	3) Fifth planets between $\sim115-130$au with decreasing masses between 1-0.04$\mathrm{M_J}$. An example intensity profile is shown in the bottom left panel of Figure \ref{fig:gamma_curves} for a fifth planet with a mass and semi-major axis of 0.63$\mathrm{M_J}$ and 120au respectively (3a in Figure \ref{fig:chi2}). No particles survive on an orbit internal to this fifth planet, however as with the previous examples, a population of particles (albeit smaller) is present in co-rotation with the fifth planet. The outer edge of the gap carved by the fifth planet therefore defines the position of the inner edge of the disk. While this fifth planet produces an intensity profile that shares a reasonable agreement with the observed profile, the slope of the inner edge is perhaps sharper than what would be expected from the observed intensity profile. For fifth planets in this region with lower masses, the gravitational potential bounding particles in co-rotation is weaker. Particles are therefore more prone to being scattered by interactions with HR8799b. We highlight this with another example fifth planet in this $\Delta\chi^2$ minimum region, with a mass and semi-major axis of 0.1$\mathrm{M_J}$ and 126au respectively (3b in Figure \ref{fig:chi2}). The intensity profile of this planet, with and without convolution with the beam of ALMA, are shown by the solid and lighter blue lines in the bottom left panel of Figure \ref{fig:gamma_curves}. Here it is clear that no particles are present in co-rotation and the inner edge of the disk is defined purely by the outer edge of the chaotic zone of the planet.
	
	4) Fifth planets with a mass and semi-major axis around 0.25$\mathrm{M_J}$ and 110au respectively. The intensity profile of this fifth planet (4 in Figure \ref{fig:chi2}) is shown in the bottom right panel of Figure \ref{fig:gamma_curves}. Here all particles are cleared internal to the orbit of the fifth planet. Moreover, strong interactions with HR8799b result in no particles being caught in co-rotation. The inner edge of the disk is therefore only defined by the outer edge of the chaotic zone of the planet. Similarly to region (3), the intensity profile from the simulation is sloped steeper at the inner edge than what is expected by the observed profile. It is also worth noting that a non-negligible percentage ($\sim40\%$) of the fifth planets in this region were found to cause an instability in the system in \S\ref{subsec:stability}. 
	
	Smaller mass planets in Figure \ref{fig:chi2} clear smaller gaps (see discussion in \S\ref{subsec:4plresult}) and take longer to do so \citep[e.g.][]{2015ApJ...799...41M, 2016MNRAS.462L.116S}. Indeed, for the lowest mass fifth planet in Figure \ref{fig:chi2} (0.01$\mathrm{M_J}$) the timescale to reach 50\% of the final surviving material fraction within the gap, from eq. (8) of \citealt{2015ApJ...799...41M} (assuming the scattering dominated regime) is $\sim$Gyr, far longer than the lifetime of HR8799. Very low mass fifth planets would therefore tend to a regime where no significant amount of material is scattered from its chaotic zone and the resulting intensity profile would be equivalent to the profile for the four known planets in isolation (Figure \ref{fig:4plintens}). Such fifth planets could therefore exist and be embedded in the outer disk around HR8799 without carving a noticeable gap.
	
	\subsection{Discussion of Results}
	\label{subsec:resultdiscuss}
	\begin{figure}
		\centering
		\includegraphics[trim={0.0cm 0cm 0cm 0cm},width=0.5\textwidth]{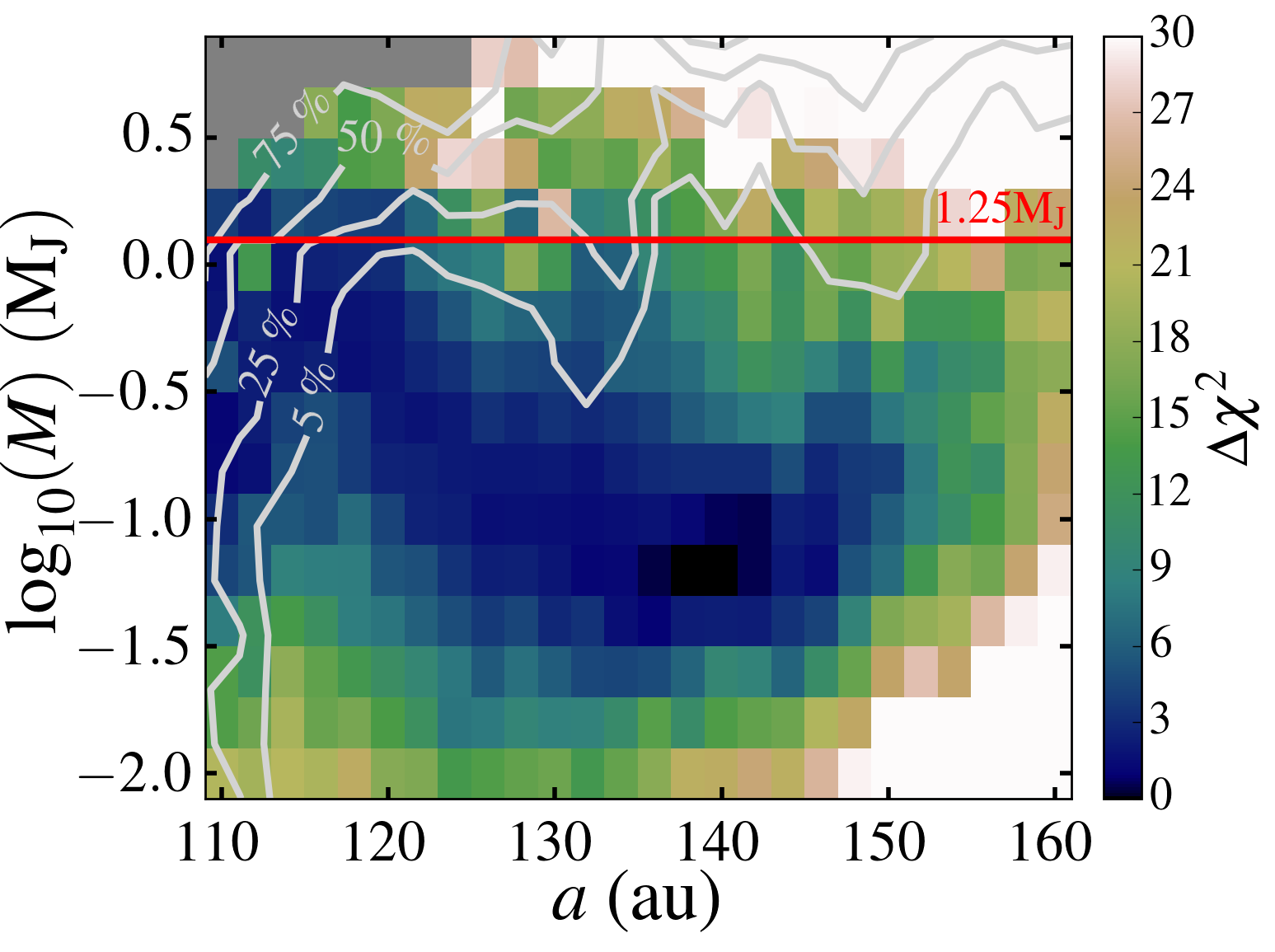}
		\caption{Identical plot to Figure \ref{fig:chi2}, however here we remove particles at the end of our simulations that have a semi-major axis within 10\% of the fifth planet and eccentricity below 0.1.}
		\label{fig:chi2wotroj}
	\end{figure}

	From Figure \ref{fig:chi2} it is clear that fifth planets in the 4 regions discussed above predict intensity profiles that agree, with different degrees of success, with the profile observed by ALMA. The profiles predicted by all these fifth planets also fit the observed profile much better compared with the four known planets in isolation, i.e. comparing Figures \ref{fig:4plintens} and \ref{fig:gamma_curves}. Indeed if the $\Delta\chi^2$ value for how well the intensity profile from the four known planets (Figure \ref{fig:4plintens}) matched the observed profile was to be plotted in Figure \ref{fig:chi2}, it would have a value of $\Delta\chi^2=60$. It would be expected therefore that the $\Delta\chi^2$ values from even lower mass planets than those simulated in Figure \ref{fig:chi2} would tend to this value, as they would not have a noticeable effect on the disk structure 
	
	All the fifth planets from the four discussed regions in Figure \ref{fig:chi2} are also below current direct imaging detection thresholds and their presence would therefore not contradict any observations. Following the original postulation from \citealt{2016MNRAS.460L..10B} therefore, we conclude that \emph{the presence of a fifth planet around HR8799 predicts an intensity profile which fits the observed profile significantly better compared with the profile predicted by the four known planets in isolation, assuming their current configuration}. Moreover we find that a fifth planet with a mass and semi-major axis of 0.1$\mathrm{M_J}$ and 138au respectively predicts an intensity profile which best fits the profile observed by ALMA.     
	
	This best fitting intensity profile contained a significant population of particles in co-rotation with the fifth planet. We therefore also consider the possibility that no particles trapped in co-rotation with the fifth planet. Indeed, the capture of material in co-rotation would depend on how the fifth planet formed, and how material was initially distributed around it. To do this, we take each of our fifth planet simulations after 60Myr and remove all particles that had a semi-major axis within 10\% of the fifth planet. Of these particles we do not remove those with eccentricities larger than 0.1. This helps stop particles being removed that are actually being scattered and happen to have a semi-major axis close to the fifth planet rather than being bound on a co-rotational orbit with the planet. 
	
	We then repeat the procedure used to create Figure \ref{fig:chi2} (e.g. performing a simultaneous $\chi^2$ minimisation over $\gamma$ and $K$ values discussed in \S\ref{subsec:getresult}) to show which fifth planet now predicts an intensity profile most like the one observed by ALMA. We show this in Figure \ref{fig:chi2wotroj}, with the colour scale again referring to $\chi^2$ minimum values relative to the overall $\chi^2$ minimum, denoted by the variable $\Delta\chi^2$. 
	
	It is clear that a fifth planet outside $\sim$150au (region (2) from \S\ref{subsec:5planresult}) no longer fits the intensity profile observed by ALMA. This is perhaps expected from the top right panel of Figure \ref{fig:gamma_curves}, as without particles in co-rotation, the gap carved by the planet is too wide such that there is not a smooth transition in the intensity profile between the edges of the gap. The $\Delta\chi^2$ values of fifth planets that were inside $\sim130$au (e.g. region (3) from \S\ref{subsec:5planresult}) are largely unchanged from values seen in Figure \ref{fig:chi2} as these did not initially show a significantly population of particles in co-rotation with the fifth planet. 
	
	Perhaps surprisingly, a fifth planet with a mass and semi-major axis of 0.063$\mathrm{M_J}$ and 140au respectively (e.g. similar to the planets in region (1) discussed in \S\ref{subsec:5planresult}) predict an intensity profile that best fits the one observed by ALMA. This shows that contributions to the predicted intensity profile from particles caught in co-rotation here are small. We note that the absolute value of the $\chi^2$ minimum does not significantly change between Figures \ref{fig:chi2} and \ref{fig:chi2wotroj}. We conclude therefore that fifth planets with a mass and semi-major axis around 0.1$\mathrm{M_J}$ and 138au respectively (region (1) from \S\ref{subsec:5planresult}) predict an intensity profile of the outer disk around HR8799 which closely agrees with the profile observed by ALMA, regardless of whether there is a significant population of material co-orbiting with the planet or not. 
	While such fifth planets are below current direct imaging detection sensitivities ($\sim1.25\mathrm{M_J}$, \citealt{2015A&A...576A.133M}) and therefore do not contradict current observations, future instruments such as the near infrared imager NIRCam available on the \textit{James Webb Space Telescope} might be able to detect such objects. Indeed, at 4.3$\mu$m NIRCam would be expected to achieve a contrast of $10^{-6}-10^{-7}$ for separations from the star $\gtrsim3''$ \citep{2010PASP..122..162B}, equivalent to $\gtrsim110$au for HR8799. For a 0.5$\mathrm{M_J}$ planet at an age of 60Myr, evolutionary models (AMES-Cond models, \citealt{2001ApJ...556..357A,2003A&A...402..701B}) suggest a planet-star contrast of 2.5$\times10^{-6}$ with the F430M NIRCam filter, highlighting that low mass planets in HR8799 would be detectable with a high SNR outside of $110$au.  
	
\begin{figure}
		\centering
		\includegraphics[trim={0.0cm 0cm 0cm 0cm},width=0.5\textwidth]{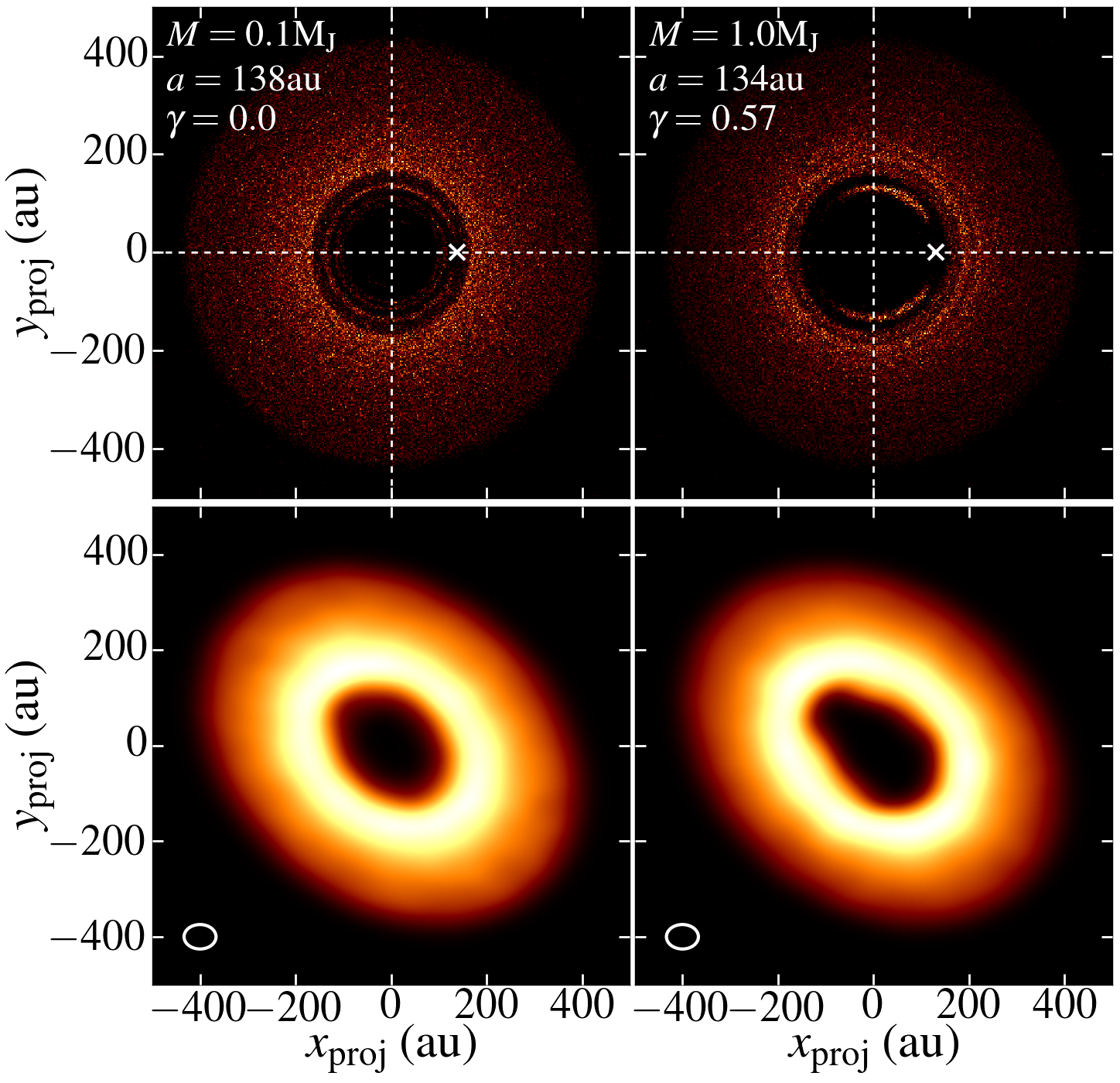}
		\caption{Intensity image of outer disk from two example fifth planet simulations without (\textit{top}) and with (\textit{bottom}) convolving with the beam size of ALMA (white ellipse). The bottom panels have an inclination and position angle derived from the ALMA observations. The last 16 intervals of the simulations were de-rotated to a frame co-rotating with the fifth planet (white cross) and stacked. The asymmetry caused by trojan particles is small and would not be expected to be detected.}
		\label{fig:stacked5thpl}
	\end{figure}

	If a significant population of particles is present in co-rotation with a fifth planet, we consider whether emission from such a population would be detectable by ALMA. We first consider the best-fit fifth planet (0.1$\mathrm{M_J}$ and 138au, top left panel of Figure \ref{fig:gamma_curves}), for which the value of $\gamma$ for this best fit was $\gamma=0.0$. The left panels of Figure \ref{fig:stacked5thpl} show the simulated intensity image including this fifth planet, without (top panel) and with (bottom panel) convolution with the ALMA beam. For consistency with the ALMA observation of the outer disk from \citealt{2016MNRAS.460L..10B}, we incline the disk in the bottom panel by $I=40^\circ$ (from face on) with a position angle of $51^\circ$ (anti-clockwise from North) before the beam convolution. Here, for this specific planet, we do not smooth each particle 1000 times around its orbit, allowing for asymmetric structure to be preserved. We find that steady state evolution is reached before 60Myr and we therefore combine the last 16 intervals from these simulations in a frame co-rotating with the fifth planet (white cross) to increase the signal to noise of the image. Particles in co-rotation with the planet librate around both the $L_4$ and $L_5$ Lagrange points in horseshoe like orbits (see \citealt{1999ssd..book.....M}). A noticeable asymmetry in the disk caused by particles co-rotating with the fifth planet, both in the images with and without convolving with the beam of ALMA, is therefore not present. 
	
	As a fifth planet becomes more massive, it would be expected that the gravitational potential bounding the $L_4$ and $L_5$ points would become deeper. For a sufficiently massive fifth planet therefore, particles in co-rotation would be expected to librate around the $L_4$ or $L_5$ points only in trojan like tadpole orbits. To highlight this, we consider a fifth planet that is similarly placed in the $\Delta\chi^2$ minimum in Figure \ref{fig:chi2} as the $0.1\mathrm{M_J}$, 138au fifth planet. We choose this planet to have a mass and semi-major axis of $1\mathrm{M_J}$ and 134au respectively (1b in Figure \ref{fig:chi2}). For reference, the intensity profile for this planet which best fit the ALMA profile is shown in the top left panel of Figure \ref{fig:gamma_curves}. The profiles with and without convolving the associated intensity images with the beam of ALMA are shown by the solid and lighter blue lines respectively. The best fitting value of $\gamma$ for this planet was also equal to $\gamma=0.57$. The intensity image itself for this fifth planet without and with convolving with the ALMA beam is shown by the top and bottom right panels in Figure \ref{fig:stacked5thpl} respectively.
	
	The top right panel of Figure \ref{fig:stacked5thpl} indeed shows that particles in co-rotation with the fifth planet are localised to the $L_4$ and $L_5$ points in trojan like orbits. However, from the bottom right panel of Figure \ref{fig:stacked5thpl} it can be seen that, the asymmetry in the ALMA convolved intensity image caused by these trojan like particles is small. Moreover if noise at the level of the ALMA observations were added to this image it would not be detectable (see Figure 1 in \citealt{2016MNRAS.460L..10B}). We conclude therefore that a significant population of particles in co-rotation with an undetected fifth planet is not ruled out by current ALMA observations. Furthermore, we conclude that our assumption of smoothing particles 1000 times around their orbit to produce Figure \ref{fig:chi2} is valid, as any asymmetric structure that would be removed in an image from using this technique would not be currently detectable by ALMA.
	
	We note that upcoming ALMA observations will provide higher resolution images of the outer disk, due to an expected beam size of 1$\times$1arcsec$^2$ (ALMA cycle 5 prog ID 2017.1.01315.S), rather than 1.7$\times$1.3arcsec$^2$ from \citealt{2016MNRAS.460L..10B}. These higher resolution images may be able to begin to disentangle the presence of material co-rotating with an additional planet, however this would depend on the overall sensitivity of the imaging.  
	
	As discussed in \S\ref{sec:HR8799}, the age of HR8799 is unclear. Based on interfermetric data \citep{2012ApJ...761...57B} and the probable membership of the Columba association \citep[e.g.][]{2014ApJ...783..121G}, a likely age of HR8799 is $\sim30-40$Myr. A younger age estimate than the 60Myr we consider during this work means that planets have less time to remove material through dynamical interactions, which may cause different populations of material to be present after 30Myr compared with 60Myr. In Appendix \ref{sec:append}, we show however that assuming HR8799 is 30Myr old rather than 60Myr has no significant effect on the results presented above. 
	
	During this work we have also assumed orbital properties of the known planets in HR8799 calculated from the work presented in \citealt{2014MNRAS.440.3140G}, notably that the masses of the four known planets are 9, 9, 9, 7$\mathrm{M_J}$ for planets e, d, c, b respectively. However, lower mass estimates also exist, with respective planet masses of, 7, 7, 7, 5$\mathrm{M_J}$, for an assumed age of 30Myr for the star \citep[e.g.][]{2010Natur.468.1080M, 2011ApJ...729..128C}. Lowering the mass of planet b from 7 to 5$\mathrm{M_J}$ has a negligible effect on how far out the chaotic zone of this planet extends to ($\sim$90au). Lowering the mass of the known planets would therefore not be expected to produce significantly different intensity profiles to the ones shown in this work, even after a fifth planet is introduced. Hence we would not expect this mass change to affect the conclusions from this work. We note however that lowering the mass of the known planets may affect the overall stability of some systems once a given fifth planet is introduced, though this would also depend on the initial orbital elements of the planets considered.    
	
	\begin{figure}
		\centering
		\includegraphics[trim={0.0cm 0cm 0cm 0cm},width=0.5\textwidth]{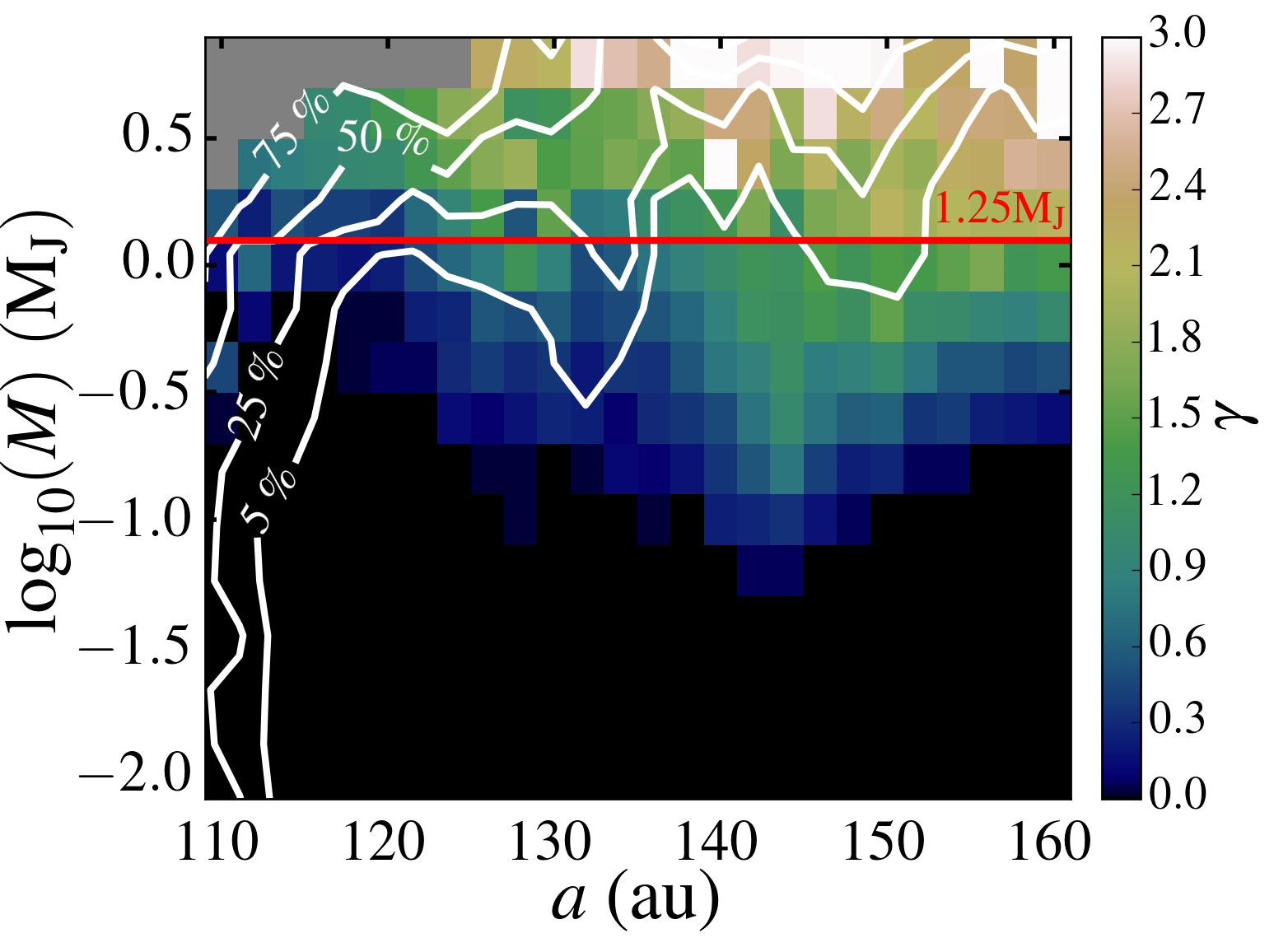}
		\caption{The values of $\gamma$ used for the intensity profiles of the outer disk for each fifth planet simulation that best fit the observed ALMA intensity profile. The initial surface density of particles in a given simulation scales with $\Sigma(r)\propto r^{-1}a_{\mathrm{ini}}^{-\gamma}$. Contours and lines are equivalent to those in Figure \ref{fig:chi2}.}
		\label{fig:gammagrid}
	\end{figure}
	
	\subsection{Surface density profiles}
	As discussed in \S\ref{subsec:getresult}, we weight the contribution of each of the particles after 60Myr in each of the fifth planet intensity images by the initial semi-major axis through the parameter $\gamma$. Larger $\gamma$ values result in particles at smaller initial semi-major axes contributing more and vice versa. We show the values of $\gamma$ which resulted in an intensity profile for each fifth planet that best fit the observed profile in Figure \ref{fig:gammagrid}. It is clear that the more massive fifth planets, the larger the best fitting value of $\gamma$. This is due to the chaotic zone around higher mass planets being larger (\S\ref{subsec:4plresult}) and therefore clearing a wider gap by the end of the simulation. With few particles surviving on orbits inwards of $\sim$150au, the ones that do survive need to be heavily weighted to significantly contribute to where the inner edge of the disk is observed to be from ALMA.   
	
	Values of $\gamma$ perhaps place constraints on the likelihood of the presence of a fifth planet, as \citealt{2016MNRAS.460L..10B} predict that the optical depth of the outer disk from the ALMA observations of HR8799 follows a $1/r$ distribution (see \S\ref{subsec:diskstruc}). As the initial particle surface density in our simulations scales roughly proportional to 1/$r^{1+\gamma}$, where $r$ is the radial distance, $\gamma$ values equal to roughly zero in Figure \ref{fig:gammagrid} would therefore be expected to be more consistent with the ALMA observations.
	
	\section{Inward delivery of particles}
	\label{sec:inwarddelv}
	For all of the simulations considered in this work, we find that the most common reason particles are removed from a given simulation is due to ejection rather than collisions with any of the massive bodies. Such a result is predicted in \citealt{2017MNRAS.464.3385W}, as large mass planets, like those around HR8799, are significantly more likely to eject particles rather than accrete them. If ejection is the dominant final outcome, we consider whether, at some point in their evolution, particles can be scattered inwards through all the planets to where the inner warm belt of material is known to be around HR8799 ($\sim$6-15au). If so, this may suggest that the inner and outer belts around HR8799 contain shared, rather than distinct populations of material. 
	
	The inner belt around HR8799 is known to have a mass of 1.1$\times$10$^{-6}$M$_\oplus$ in small grains ($\sim$1.5-4.5$\mu$m, \citealt{2009ApJ...705..314S}). The total mass however, could be significantly larger once grains of larger sizes are included. Collisional evolution between large grains creates a characteristic size distribution, with the smallest grains being ejected from the system due to radiation pressure. Assuming a characteristic size distribution that scales as $n(D)\propto D^{2-3q}$, where $q=11/6$ \citep[e.g.][]{2007ApJ...658..569W,2007ApJ...663..365W}, the mass loss rate in $\mathrm{M}_\oplus$/yr due to the ejection of small particles can be estimated by 
	\begin{equation}
	\frac{dM_\mathrm{loss}}{dt}=1700f_\mathrm{obs}^2r_\mathrm{disk}^{0.5}L_\star M_\star^{-0.5}\left(\frac{r_\mathrm{disk}}{dr_\mathrm{disk}}\right),
	\label{eq:massloss}
	\end{equation}
	where $f_\mathrm{obs}$ is the fractional luminosity of the disk, $r_\mathrm{disk}$ is the radial position of the disk in au, $dr_\mathrm{disk}$ is the radial width of the disk in au, $L_\star$ is the luminosity of the star in $L_\odot$ and $M_\star$ is the mass of the star in $M_\odot$ (see eq. (29) in \citealt{2007ApJ...658..569W}). We note that this mass loss rate is not dependent on the maximum grain size and therefore mass of the disk. For the inner disk around HR8799, $f_\mathrm{obs}=2.2\times10^{-5}$, $r_\mathrm{disk}=10$au and $dr_\mathrm{disk}=9$au \citep{2009ApJ...705..314S}, giving a mass loss rate from eq. (\ref{eq:massloss}) of $dM_\mathrm{loss}/dt=1.2\times10^{-5}\mathrm{M_\oplus/Myr}$.
	
	Mass input into the inner disk can come from embedded bodies with sufficiently large sizes that they take longer than the lifetime of the system to collide (i.e. an Asteroid belt). However, here we consider that mass may also be replenished due to input from an external source, that is, planetesimals that are scattered inward from the outer belt.
	
	We consider the five previously described simulations from \S\ref{subsec:5planresult}, which contained a fifth planet around HR8799 which produced intensity curves that agreed well with the curve observed by ALMA. That is, the fifth planets with masses and semi-major axes of 0.1$\mathrm{M_J}$ at 138au, 0.25$\mathrm{M_J}$ at 148au, 0.63$\mathrm{M_J}$ at 120au, $0.1\mathrm{M_J}$ at 126au and $0.25\mathrm{M_J}$ at 110au respectively (we note that we do not consider the fifth planet with 1$\mathrm{M_J}$ at 134au discussed in \S\ref{subsec:resultdiscuss}). We repeated these simulations, but this time removed the massless particles from the simulations as soon as any part of their orbit reached within 10au. This represents the distance where we deemed particles to have 'joined' the inner disk. Over a given time in these simulations therefore, the rate at which particles join the inner disk could be obtained. 
	
\begin{figure}
		\centering
		\includegraphics[trim={0.0cm 0cm 0cm 0cm},width=0.5\textwidth]{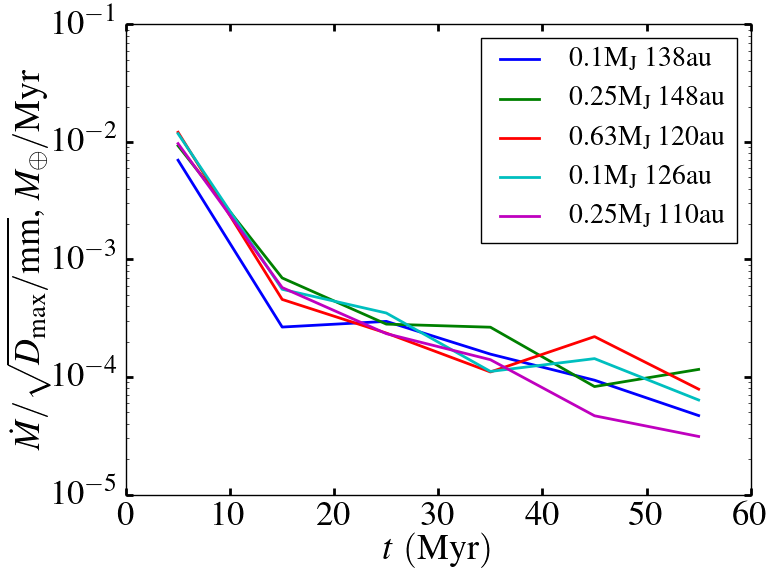}
		\caption{Rate at which mass reaches the inner disk due to inward scattering of particles from the outer disk by the planets. Bin sizes for time are 10Myr. To get the mass input rate into the inner disk, it is required to multiply by an efficiency factor discussed in text. $D_\mathrm{max}$ is the maximum grain size considered in mm. The rate at which mass reaches the inner disk does not significantly change for different fifth planet masses and semi-major axes.}
		\label{fig:inputrate}
	\end{figure}

	To convert this inward flux of particles into a rate of inward delivery of mass, each particle was assigned mass by scaling to the current mass of the outer belt around HR8799 in 10-1000$\mu$m grain sizes of 1.2$\times10^{-1}$M$_\oplus$ \citep{2009ApJ...705..314S}. Since there are likely also objects larger than 1mm in the disk we assume that the largest objects (in both the disk and in the scattered material) have a size $D_\mathrm{max}$ resulting in a total mass input rate from a given simulation, $\dot{M}$, during successive 10Myr epochs, as shown in Figure \ref{fig:inputrate}, where this mass input scales with the square root of the maximum grain size, $D_\mathrm{max}$ (see eq. (3) in \citealt{2007ApJ...658..569W}).       
	
	Figure \ref{fig:inputrate} shows that despite the different masses and semi-major axes of the fifth planet considered, the mass input to the inner disk from the outer disk for all epochs is largely unchanged. The mass input during the final 10Myr of the simulations implies an input rate of 5$\times10^{-6}$($D_\mathrm{max}/\mathrm{mm}$)$^{1/2}$M$_\oplus$/Myr. For $D_\mathrm{max}\sim1$mm this mass input rate is roughly equivalent to the mass loss rate calculated from eq. (\ref{eq:massloss}), suggesting that the mass loss rate from the inner disk can be replenished by 10-1000$\mu$m grains scattered inward from the outer belt. However, first we must consider whether all the grains that are scattered in from the outer belt and reach 10au actually get incorporated into the inner disk or whether they simply pass through the belt to get ejected at a later time. 
	
	If we assume that it is collisions that allow material to be incorporated into the inner disk, then the efficiency of mass input rate into the inner disk can be estimated by comparing the timescale for which grains would be expected to mutually collide, $t_\mathrm{col}$, with the timescale for which grains would be expected to be ejected by interactions with the innermost known planet HR8799e, $t_\mathrm{scat}$. That is, the efficiency is equivalent to
	\begin{equation}
	\eta_\mathrm{eff}=\frac{t_\mathrm{scat}}{t_\mathrm{col}}.
	\label{eq:eff}
	\end{equation}
	The ejection timescale, $t_\mathrm{scat}$ is given by eq. (2) in \citealt{2017MNRAS.464.3385W}
	\begin{equation}
	t_\mathrm{scat}=10^9\left(\frac{M_\star^{3/4}a^{3/4}}{M}\right)^2,
	\label{eq:tscat}
	\end{equation}
	in yr, where $a$ and $M$ are the semi-major axis and mass of a planet in au and $\mathrm{M_\oplus}$ respectively and $M_\star$ is the mass of the star in $\mathrm{M_\odot}$. For HR8799e, $t_\mathrm{scat}$ is equal to 14kyr. Combining eq. (6), (13) and (25) from \citealt{2007ApJ...658..569W}, the collisional timescale of objects of size $D_\mathrm{max}$ in mm at the inner disk is
	\begin{equation}
	t_\mathrm{col}(D_\mathrm{max})=0.04r_\mathrm{disk}^{1.5}\left(\frac{dr_\mathrm{disk}}{r_\mathrm{disk}}\right)\frac{1}{f_\mathrm{obs}}\left(\frac{10^3D_\mathrm{max}}{0.8L_\star\left(\frac{2700}{\rho}\right)}\right)^{1/2},
	\label{eq:tcolany}
	\end{equation}
	where $\rho$ is the density of particles in kg m$^{-3}$. Therefore, we infer an efficiency of mass input rate that scales as $\eta_\mathrm{eff}\propto D_\mathrm{max}^{-1/2}$. Since the rate at which mass reaches the inner disk shown in Figure \ref{fig:inputrate} scales $\propto D_\mathrm{max}^{1/2}$, we find that the rate at which mass is incorporated into the inner disk is insensitive to the largest particle size. That is, while larger maximum grain sizes increase the amount of mass that is scattered inward to reach the inner disk, the collisional timescale for these grains also increases, such that the amount of mass that is actually incorporated into the inner disk remains constant. 
	
	For an assumed grain density of 2700kg $\mathrm{m^{-3}}$, we find the mass input rate into the inner disk from inward scattering from the outer disk, including the efficiency factor given by eq. (\ref{eq:eff}), to be 8.8$\times$10$^{-8}\mathrm{M_\oplus/Myr}$, equal to $\sim1\%$ of the mass loss rate. We conclude therefore that the mass loss from the inner belt cannot be replenished by inward scattering of material from the outer disk for the current age of HR8799, with this conclusion being independent of the maximum grain size of inward scattered material. The inner belt around HR8799 is therefore more likely to be a distinct population of material, akin to the Asteroid belt in the Solar System, rather than containing a significant amount of material from the outer belt. 
	
	We note however that the mass input rate is $\sim$100 times larger in Figure \ref{fig:inputrate} during the first 10Myr. This suggests that if HR8799 were observed in the first 10Myr of its lifetime, the warm dust emission from the inner belt due to material being scattered inwards from the outer belt would be comparable in brightness to that observed towards HR8799 today. This perhaps has implications for other young systems which have been observed to host belts of particularly hot and warm dust (e.g. $\eta$ Telescopii, HD95086, \citealt{2009A&A...493..299S, 2015ApJ...799..146S}), as brighter than expected emission from an inner belt could be an indication of inward scattering of material from outer planets interacting with an outer belt of material (\citealt{2012MNRAS.420.2990B, 2012A&A...548A.104B}; Marino et al. submitted).     
	
	\section{Summary and Conclusions}
	\label{sec:conc}
	
	We simulated how an outer population of material interacts with the four known planets around HR8799. We found that the intensity profile of the outer disk that is formed in our simulations does not agree with the equivalent intensity profile observed by recent ALMA observations \citep{2016MNRAS.460L..10B}. Notably the inner edge of the disk from our simulations was much further in than that suggested by ALMA. We therefore support the postulation in \citealt{2016MNRAS.460L..10B} that the four known planets in their current configuration do not sculpt an outer disk that is consistent with the one that is currently observed.  
	
	We subsequently added an additional fifth planet with a range of masses and semi-major axes in our simulations that was external to the outermost known planet. We found that a fifth planet with a mass and semi-major axis around 0.1$\mathrm{M_J}$ and 138au respectively produced an outer disk in our simulations with an intensity curve that best fit the curve observed by ALMA. However, fifth planets with semi-major axes between 140-160au for increasing masses between 0.04-1$\mathrm{M_J}$, between 115-130au with decreasing masses between 1-0.04$\mathrm{M_J}$ and fifth planets with a mass and semi-major axis around 0.25$\mathrm{M_J}$ and 110au respectively also predict an intensity curve for the outer disk that agrees well with the one observed with ALMA. Moreover, we found that these fifth planets can remain dynamically stable with the known planets for the lifetime of the system and are below the current sensitivity threshold for detection via direct imaging surveys. We conclude therefore that the presence of a fifth planet around HR8799 that is external to the outermost known planet, predicts an intensity profile that fits the one observed significantly better compared with the profile predicted by the four known planets in isolation, assuming their current configuration. 
	
	In order for many of the simulated fifth planets to produce a well fitting intensity profile for the outer disk, material needed to be present in co-rotation with the fifth planet. However, after artificially removing material in co-rotation with the fifth planet, we found that a fifth planet with a mass and semi-major axis around 0.1$\mathrm{M_J}$ and 138au respectively still predicted an intensity profile that best fit the profile observed with ALMA. We therefore concluded that, regardless of whether material in co-rotation with a fifth planet is considered or not, the predicted intensity profile fits the observed profile much better than the profile predicted by the four known planets in isolation. Moreover, we found that any asymmetric structure in the outer disk indicative of material in co-rotation with a fifth planet would not be detectable in the current ALMA observations.
	
	We also considered whether a significant amount of material could be passed from the outer disk around HR8799 through the known planets and an additional fifth planet to the inner disk. That is, whether mass loss of the smallest grains in the size distribution of the inner disk, due to radiation pressure, could be replenished by inward scattering of material from the outer belt. We found that the amount of material that is scattered through the planets from the outer to the inner belts does not significantly change for different masses and semi-major axes of the additional fifth planet. We assumed that the efficiency of mass input into the inner disk from inward scattering is equal to the ratio between the timescale for inwardly scattered material to collide at the inner belt and the timescale for their ejection by interactions with HR8799e. We found that only $\sim1\%$ of the mass loss rate of the inner disk can be replenished by inward scattering of material from the outer belt. This result is independent of the considered grain size of inward scattered material. We conclude therefore that the inner disk around HR8799 is most likely a distinct population of material akin to the Asteroid belt in the Solar System. However we find that if HR8799 were observed in the first 10Myr of its lifetime, the emission from the hot dust at the inner belt would be expected to be similar in brightness to the hot dust seen around HR8799 today, due to a larger amount of material being scattered inwards from the outer belt. This perhaps has implications for young systems with particularly bright populations of hot and warm dust (e.g. $\eta$ Telescopii, \citealt{2009A&A...493..299S}), as this may be indicative of the inward scattering of material from an outer disk by planetary objects.

	\section*{Acknowledgements}
	We greatly thank Mark Booth for providing us with the ALMA data for HR8799 presented in \citealt{2016MNRAS.460L..10B}, as without this, much of the analysis discussed in this paper would not be possible. We also thank the reviewer for their insightful comments, which lead to the improvement of this manuscript.
	
	Simulations in this paper made use of the REBOUND code which can be downloaded freely at http://github.com/hannorein/rebound.
	
	MJR acknowledges support of an STFC studentship and MCW acknowledges the support from the European Union through grant number 279973. GMK acknowledges support by the Royal Society as a Royal Society University Research Fellow.

	
	
	
	\bibliographystyle{mnras}
	\bibliography{example} 

	
	
	\appendix
	\section{A younger age estimate for HR8799}
	\label{sec:append}
	\begin{figure}
		\centering
		\includegraphics[width=0.5\textwidth]{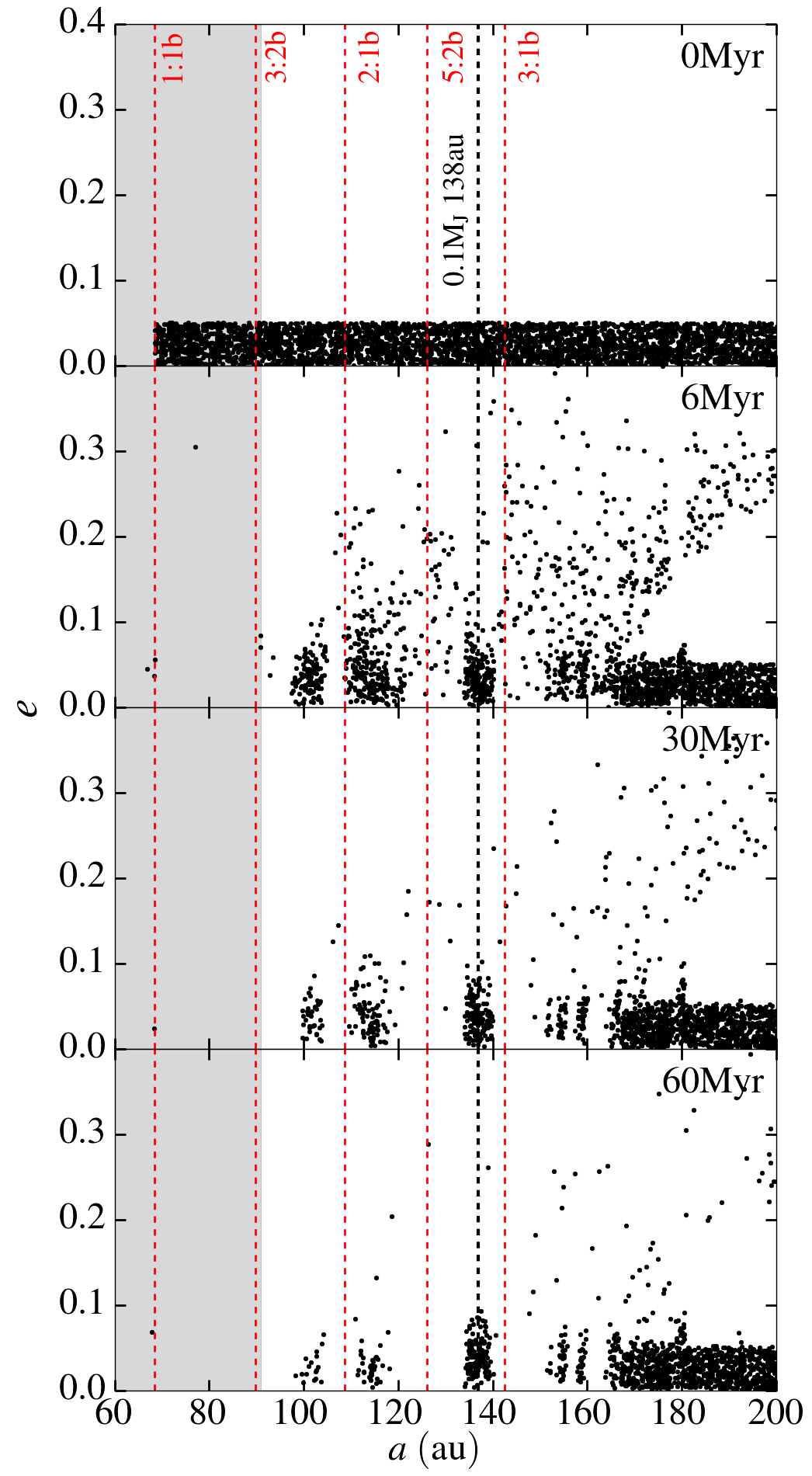}
		\caption{Semi-major axis vs. eccentricity of particles evolving due to dynamical interactions with the four known planets and an additional fifth planet with a mass and semi-major axis of 0.1$\mathrm{M_J}$ and 138au respectively. This fifth planet predicts an intensity profile which agrees most with the ALMA observed profile (see Figure \ref{fig:chi2}). The red dashed lines refer to 1st order MMRs of HR8799b and the black dashed line refers to the semi-major axis of the fifth planet.}
		\label{fig:eplotfifthplanet}
	\end{figure}
\begin{figure}
		\centering
		\includegraphics[width=0.5\textwidth]{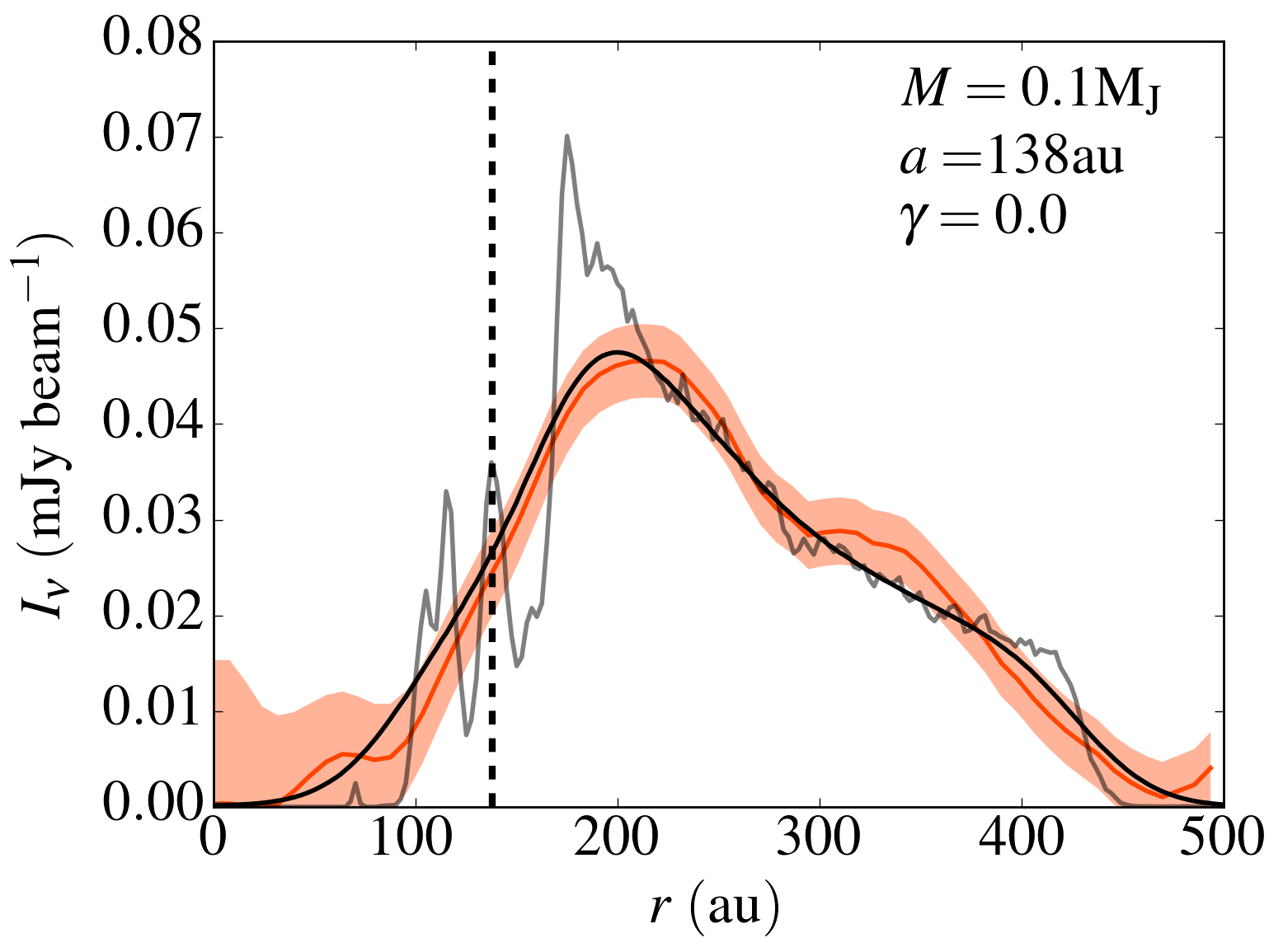}
		\caption{Comparison of intensity profile generated by a fifth planet with a mass and semi-major axis of 0.1$\mathrm{M_J}$ and 138au respectively after 30Myr, with and without convolving with the beam of ALMA (black and grey lines respectively), with the profile observed by ALMA (orange line). The orange shaded region gives the 1$\sigma$ rms of the noise per beam. The semi-major axis of the fifth planet is given by the dashed vertical line. The value of $\gamma=0$ refers to an initial scaling of the surface density of particles of $\Sigma(r)\propto r^{-1}$.}
		\label{fig:panel}
	\end{figure}

	Here we consider how an age estimate of HR8799 of 30Myr rather than 60Myr affects the results presented in \S\ref{subsec:5planresult}. We initially show the semi-major axis vs. eccentricity distribution of particles interacting with the four known planets and an additional planet with a mass and semi-major axis of 0.1$\mathrm{M_J}$ and 138au respectively in Figure \ref{fig:eplotfifthplanet}. We note that the simulated outer disk produced with this planet had an intensity profile that agreed most with the ALMA observed profile as described in \S\ref{subsec:getresult}. All lines and shaded regions are identical to those in Figure \ref{fig:evsa}, with the semi-major axis of the fifth planet given by the black dashed line. Figure \ref{fig:eplotfifthplanet} shows that few particles are scattered after 30Myr. Indeed, the intensity profile predicted after 30Myr (shown in Figure \ref{fig:panel}, with all lines taking identical definitions to those described in Figure \ref{fig:gamma_curves}), is not significantly different from the profile produced after 60Myr (top left panel of Figure \ref{fig:gamma_curves}), highlighting that an assumed age of 30Myr rather than 60Myr for HR8799 does not significantly affect how well the intensity profile generated by this fifth planet agrees with the ALMA observed profile. 
	
\begin{figure}
		\centering
		\includegraphics[width=0.5\textwidth]{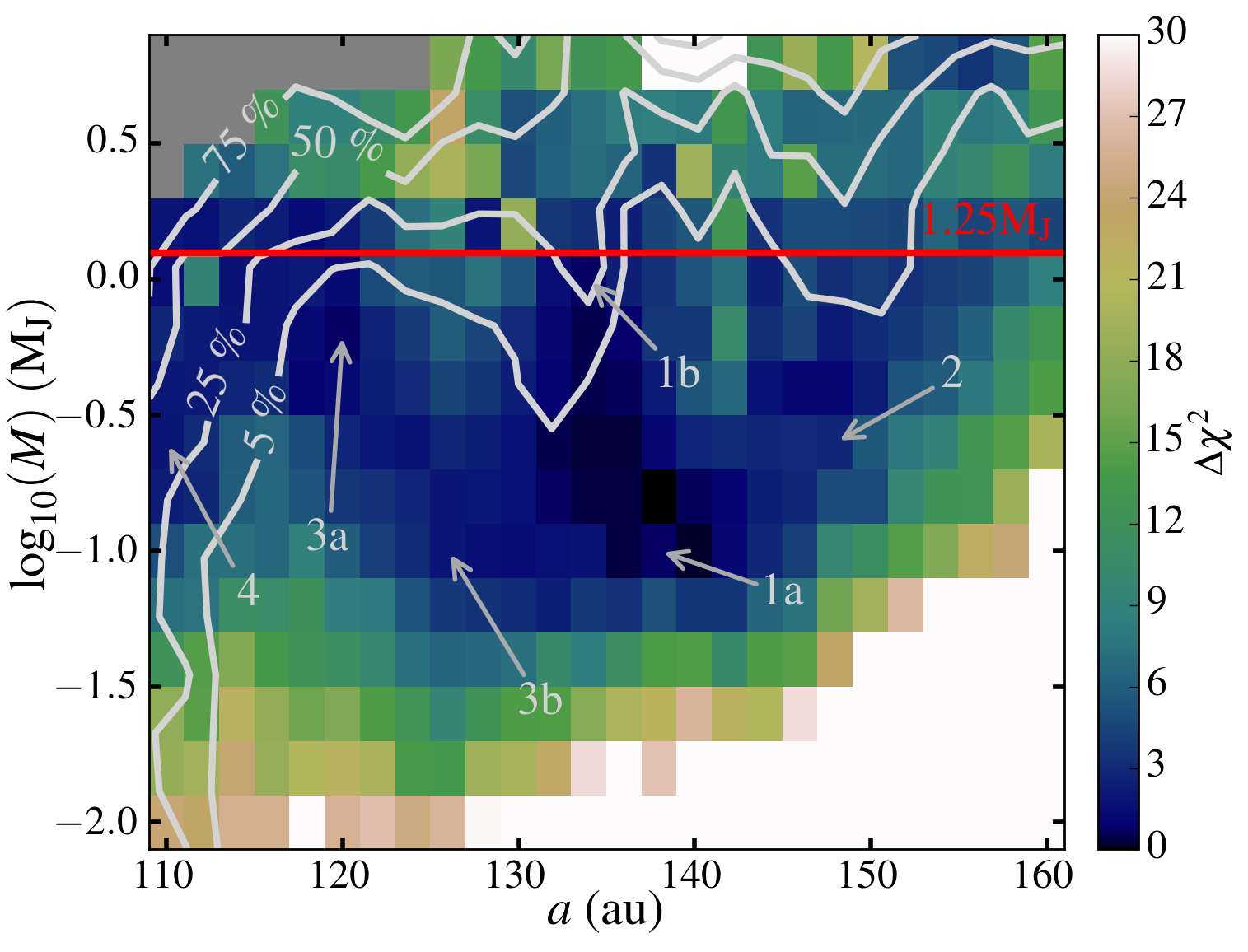}
		\caption{Identical plot to Figure \ref{fig:chi2}, however here simulations are run to 30Myr rather than 60Myr. Annotations and contours are identical to those shown in Figure \ref{fig:chi2} for reference. Running simulations to 30Myr rather than 60Myr has little effect on the types of fifth planets which predict intensity profiles that well fit the profile observed by ALMA.}
		\label{fig:incpl}
	\end{figure}

	More generally, we also investigated how an assumed age of 30Myr rather than 60Myr affects the intensity profiles produced by all the fifth planets considered in this work. We repeat the method described in \S\ref{subsec:getresult} to produce Figure \ref{fig:chi2}, that is, calculating $\Delta \chi^2$ values for all fifth planets, which describes how well the simulated intensity profile agrees with the ALMA observed profile. However now we run simulations to 30Myr rather than 60Myr. Figure \ref{fig:incpl} shows these $\Delta \chi^2$ values, with all contours and annotations being identical to those given in Figure \ref{fig:chi2} for reference. Comparing Figures \ref{fig:chi2} and \ref{fig:incpl} it is clear that assuming an age of 30Myr rather than 60Myr has little effect on the types of fifth planets which produce intensity curves that strongly agree with the ALMA observed profile. We also note that the absolute $\chi^2$ values change little between Figures \ref{fig:chi2} and \ref{fig:incpl}.   
	

	\bsp	
	\label{lastpage}
	
\end{document}